# EUTelescope Guide for ATLAS ITk Strip: Reconstruction & Analysis

Daniel Rodriguez (IFIC) & Edoardo Rossi (DESY)

Version 1.2.1

**Abstract**

EUTelescope is arguably the most used software to reconstruct tracks from telescope data. This guide explains how to install a version adapted for the reconstruction of ITk Strip test beam data. In this version, an example of such reconstruction with 2016 data is provided. In addition, this example will be explained step-by-step and a general overview of how to operate EUTelescope is given. It is also briefly explained how the post-reconstruction analysis has been performed.
The aim of this guide is to be used a starting point for EUTelescope beginners in future test beam reconstruction and analysis works. It is focused on reconstruction for the ATLAS ITk Strip detector using a binary read-out.

# Contents

# 1 Introduction

In this guide, the basic knowledge about the reconstruction and the analysis for the ATLAS ITk Strip is given. This is not an official document. The guide is aimed at beginners who are starting using EUTelescope. The first section contains a very quick overview of the framework. In the second section, it is explained how to install EUTelescope. Then, in the third section the basics of EUTelescope are explained using some hands-on examples that run out-of-the-box. Finally, in the fourth section it is illustrated how to run the post-reconstruction analysis. All these steps are aimed for members of the ATLAS ITk Strip test beam community that want to analyse data from a binary read-out.
This guide is not aimed to be an advanced guide. Much more knowledge should be acquired to become proficient with EUTelescope. Unfortunately, most of this knowledge can be acquired only overtime by working with EUTelescope.
Moreover, some extra information has been included using **Kenny Wraight's** notes.

## 1.1 Introduction to EUTelescope

The EUDET-type telescopes consist in six planes with high-resolution MIMOSA26 pixel sensors. They are built to provide a very flexible setup for measurements with a Device Under Test (DUT). The EUDET-type telescopes have a very flexible setup for measurements with a Device Under Test (DUT), thanks to their hardware, read-out electronics and Data Acquisition system (EUDAQ).

The EUTelescope framework has a modular structure (as shown in Figure 1.1) and it is a group of Marlin (Modular Analysis & Reconstruction for the LINear collider) processors. It has been created in the context of the EUDET project to reconstruct tracks obtained with pixel beam telescopes. The tracks obtained by the reconstruction are used to characterize a DUT, such as strip sensor, or measurements of scattering length of passive materials. EUTelescope is embedded in the ILCsoft framework. The core elements of this framework are the Linear Collider I/O (LCIO) data format, the Geometry API for Reconstruction (GEAR) geometry description toolkit, and Marlin.

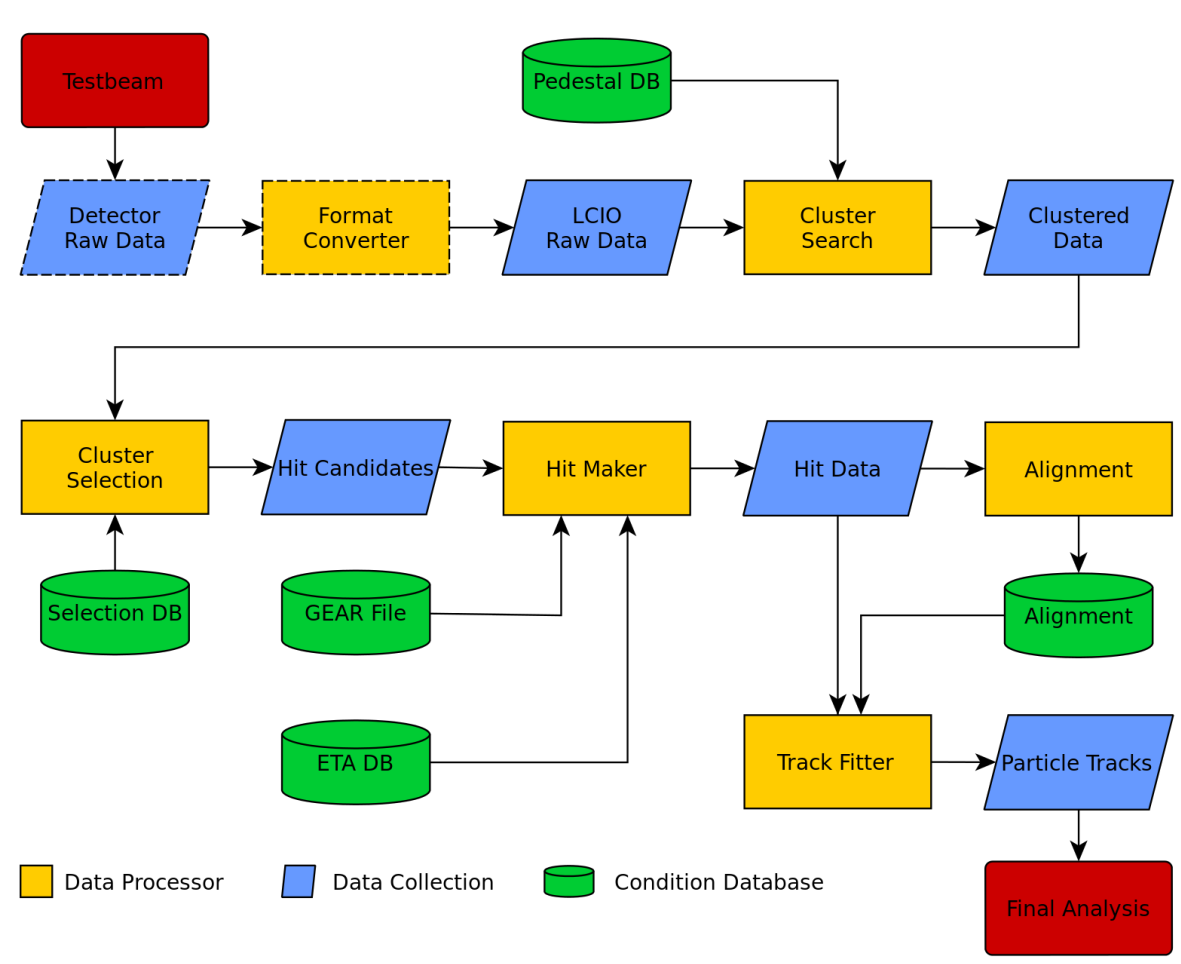

Figure 1.1: EUTelescope framework

## 1.2 External Links

This section is going to be a compilation of links.
The main EUTelescope web:

*http://eutelescope.web.cern.ch*

Where the important information that you can find about:

LCIO Data Format:

*http://eutelescope.web.cern.ch/content/lcio-data-format*

Gear File (Telescope geometry description):

*http://eutelescope.web.cern.ch/content/gear-telescope-geometry-description*

RAIDA Processor:

*http://eutelescope.web.cern.ch/content/raida-processor*

Concerning EUDAQ:

*https://telescopes.desy.de/EUDAQ*

EUDAQ User Manual:

*http://eudaq.github.io/manual/EUDAQUserManual.pdf*

Moreover, more information regarding the EUDAQ installation:

*http://eudaq.github.io*

Finally, some other guides with useful information:
EUTelescope 1.0: Reconstruction Software for the AIDA Testbeam Telescope:

*https://cds.cern.ch/record/2000969/files/AIDA-NOTE-2015-009.pdf*

Performance of the EUDET-type beam telescopes

*https://arxiv.org/abs/1603.09669*

Documentation about GBL track fitting

*https://github.com/AlexanderMorton/GBLManuel/blob/master/thesis.pdf*

# 2 Installation

This section explains how to install EUTelescope on **lxplus** or **NAF**. The aim is to provide a very basic guide even with people that have never worked with this.

## 2.1 Prerequisites and Compilation

First of all, enter to lxplus[1]:

*ssh -Y yourusername@lxplus.cern.ch*

After entering to lxplus, we can check that the shell is *bash*, due to sometimes the native shell is *tsch*:

*echo $SHELL*

The result should be:

*/bin/bash*

If the it is not the result, we should use the following command in order to change to *bash* shell[2]:

*bash -l*

Then, we can continue loading the ATLAS setup, and the right versions of cmake and gcc:

*setupATLAS*
*lsetup "gcc gcc493_x86_64_slc6"*
*lsetup "cmake 3.3.2"*

If you are working on NAF, the command *setupATLAS* must be replaced by:

*export ATLAS_LOCAL_ROOT_BASE=/cvmfs/atlas.cern.ch/repo/ATLASLocalRootBase*
*export DQ2_LOCAL_SITE_ID=DESY-HH_SCRATCHDISK*
*source /cvmfs/atlas.cern.ch/repo/ATLASLocalRootBase/user/atlasLocalSetup.sh*

It is suggested to always use the latest version available of cmake and gcc[3]. To check which is the most recent available version use:

*lsetup gcc -h lsetup cmake -h*

Now, you have to create the folder where you want to install EUTelescope and define it as the path $ILCSOFT:

*export ILCSOFT="/path/where/you/want/to/install/EUTelescope"*
*mkdir -p $ILCSOFT*
*cd $ILCSOFT*

And then download from svn the folder with the files needed for the installation:

*svn co https://svnsrv.desy.de/public/ilctools/ilcinstall/branches/v01-17-10-eutel-pre/ ilcinstall_v01-17-10-eutel*

[1]Along the guide, it has been used *user/r/yourusername* where *r* is the first letter of the user name used
[2]More information in http://www.gnu.org/software/bash/manual/bashref.html
[3]For this guide, the versions used are the previous ones shown.

Before running the compilation, one must change the root version that will be downloaded and used by the EUTelescope compilation. For the time being, the compilation does not work with the present choice of ROOT 6. To change the root version, change in the file **ilcinstall_eutel-git/v01-17-10/release-versions.py** at line 145, the version to be used if cpp11 is used (the standard choice):

*gedit ilcinstall_v01-17-10-eutel/releases/v01-17-10/release-versions.py*

This is how it should appear after the modification (in bold letters):

*if use_cpp11:*
*ROOT_version = "**5.34.30**"*
*print " I install ROOT version : 6. " , ROOT_version , use_cpp11*
*else:*
*ROOT_version = "5.34.30"*
*print " I install ROOT version : 5. " , ROOT_version , use_cpp11*

Of course, it would be equivalent to just remove the if-else statements, and install ROOT 5.34.30 in any case.
Now, one can **run the compilation**:

*cd $ILCSOFT/ilcinstall_v01-17-10-eutel*
*./ilcsoft-install -i examples/eutelescope/release-standalone.cfg*

At this point take a break, and after a couple of hours the compilation should be finished. Note that if you have a problem at this point, and you want to re-run the compilation, you should delete the folder were EUTelescope was being installed, i.e. the folder *$ILCSOFT/v01-17-10*.
Now you have installed and compiled EUTelescope and all the programs needed to make it run. These are installed in the folder *$ILCSOFT/v01-17-10*, that should contain the following files and folders:

*CED Eutelescope ILCSoft.cmake init_ilcsoft.sh Marlin root CLHEP GBL ILCSoft.cmake.env.sh lccd MarlinUtil CMake gear ilcutil lcio RAIDA*

Now, the last tagged EUTelescope version is installed (as of March 2017, it is the version v1.0). This guide is written for another version which includes the GBL alignment and tracking. EUTelescope is automatically installed as a git repository, thus it is possible to checkout any other **branch** using git. To do this, enter in the EUTelescope folder (*$ILCSOFT/v01-17-10/Eutelescope/v1.0*) and use the following commands:

*git remote add Edo https://github.com/edorossi/eutelescope.git*
*git fetch Edo*
*git checkout -b EdoBranch Edo/ITkStripExample*

If everything worked well, the last output in the terminal will be:

*Switched to a new branch 'EdoBranch'*

Now, one needs to setup the EUTelescope environment. In order to do this, the following files must be sourced:

*source $ILCSOFT/v01-17-10/ILCSoft.cmake.env.sh*
*source $ILCSOFT/v01-17-10/init_ilcsoft.sh*
*source $ILCSOFT/v01-17-10/Eutelescope/v1.0/build_env.sh*

The installation provides the last stable EUDAQ version. EUDAQ is the data acquisition code of the telescope and it creates the raw data, the starting point of the reconstruction. Then, the format of the raw data depends on the **EUDAQ version**. For this reason, EUDAQ must be compiled inside EUTelescope, in order to be able to convert the raw data in the LCIO format. The best choice is to set-up in EUTelescope the same version of EUDAQ that was used in the data acquisition. EUDAQ can be found in *$ILCSOFT/v01-17-10/Eutelescope/v1.0/external/eudaq* and the version is the name of the folder.
To download EUDAQ v1.6-dev as used in the 2016 test-beams, one can use the git commands:

> *git clone https://github.com/eudaq/eudaq.git*
> *git checkout tags/v1.6-dev-end*

You can simply rename the new folder as the original EUDAQ folder. In this way, EUTelescope will found the new EUDAQ in place of the original one, and it is not necessary to modify the source files.

> *mv v1.6.0 v1.6.0_original*
> *mv v1.6-dev-end v1.6.0*

Pay attention, the best practice consists of not deleting the original folder, but simply rename it. In this way, it will be still available in case the new version creates problems. Another, more elegant, option is not to rename the new directory, and change the source file *build$_e$nv.sh* in order to link EUTelescope to the right EUDAQ folder.
At this point, the new EUDAQ must be built (note that this must be done even if you changed version with git). To do this, enter into */afs/cern.ch/user/r/yourusername/your/path/to/v01-17-10/Eutelescope/v1.0/external/eudaq/v1.6.0/build*. EUDAQ can be built only with gcc 4.8 or more. It has been noticed that often at this point the version of gcc in use is changed. You can check it with:

> *gcc --version*

In case the version is not enough, you should switch to a newer one, for example gcc 4.9:

> *lsetup "gcc gcc493_x86_64_slc6"*

Again, the best practice is to always use the latest version available.
Then, enter the build folder, and build EUDAQ activating the flag for the Native Reader (needed for the conversion of the raw data):

> *cmake -D BUILD_nreader=ON ..*
> *make install -j*

If you have problems, it is probably due to the fact that EUDAQ can not built some parts that are actually not used for the reconstruction. To check which flags are activated, use:

> *cmake -L*

And turn off all the flags apart from the Native Reader (it has been noticed that normally the problematic flags are main, onlinemon and gui), for example:

> *cmake -D BUILD_nreader=ON -D BUILD_main=OFF -D BUILD_gui=OFF -D BUILD_onlinemon=OFF ..*
> *make install -j*

Another problem that has been noticed is related to the *-j* of make install. If the compilation does not work at this point, try to use *-j8* instead of (limiting the use of the CPU to 8 core), or only *make install*.
If EUDAQ compiles well, it is necessary to do again:

> *export MARLIN_DLL=/afs/cern.ch/user/r/yourusername/your/path/to/v01-17-10/Eutelescope/v1.0/external/eudaq/v1.6.0/lib/libNativeReader.so*

Finally, before starting with some example, it is necessary to install all libraries. So, inside of build folder, */afs/cern.ch/user/r/yourusername/your/path/to/v01-17-10/Eutelescope/v1.0/build*, you have to delete everything and after that, installing it again. That is because not all libraries were loaded. With the new installation, you are loading all libraries. So inside the build folder (empty in this moment), execute the next commands:

> *cmake ..*
> *make install*

At this point, EUTelescope should be setup correctly and it should be ready to be used. Pay attention, that for every fresh terminal it is necessary to source the various files, like explained before, and export the variable MARLIN_DLL. It is very useful to include these instructions in your .bashrc file.

#### 2.1.1 EUDAQ 2.0

In case data was taken using EUDAQ 2.0, the conversion from raw data to the lcio format is done by EUDAQ alone, without the need of the conversion been called by EUTelescope. In such a case, one does not need to install the right EUDAQ version inside EUTelescope, but it is suggested to install EUDAQ v1.6-dev-end (as explained in the previous sections) in order to have retro-compatibility with older data.
First, download the right EUDAQ v2.0 branch:

> *git clone https://github.com/eyiliu/eudaq.git*
> *cd eudaq*
> *git checkout itkstrip*

Then, before compiling it, one needs to fix some conversion parameters to make it compatible with 2017 data. First, to overcome the overlap of the planes' ID between the FEI4 and the DUT, the ID of the former must be changed. To do this, open *user/eudet/module/src/UsbpixrefRawEvent2LCEventConverter.cc* and modify the line:

> *uint32_t chip_id_offset = 10;*

to:

> *uint32_t chip_id_offset = 7;*

In this way, the FEI4 plane will be assigned to ID 7.
If you want to manipulate the strip numbers [4] open the file *user/itkstrip/module/src/ItsAbcRawEvent2LCEventConverter.cc* and change the lines:

> *zsFrame->chargeValues().push_back(i);//x*

[4] This is needed, for example, for R0 modules

This line assign the strip number and it is possible to modify it depending on what is needed. send me more information or I do After this modifications, EUDAQ can be built. It is suggested to use gcc 6.2.0 (*lsetup "gcc gcc620_x86_64_slc6"*), cmake 3.6.0 (*lsetup "cmake 3.6.0"*) and root 6 (*localSetupROOT*). Create a *build* folder and then use the commands:

*cmake -D EUDAQ_BUILD_STDEVENT_MONITOR=OFF -D USER_ITKSTRIP_BUILD_ITSROOT=OFF ..*
*make install -j8*

Once EUDAQ is built, the converter executable is *bin/euCliConverter* and can be used with:

*./euCliConverter -i /path/to/input/file/runXXXXXX.raw -o /path/to/output/file/runXXXXXX.slcio -ip*

For the output name , it is suggested to use the format *runXXXXXX.slcio* independently from the input file name. This slcio file will be used as the input for the reconstruction.
The example *ITkEUDAQ2* contains the steering files for EUDAQ v2.0. The only difference with respect of the reconstruction with EUDAQ v1.6 is that the first step name *noisypixel* instead of *converter*.

## 2.2 .bashrc proposed

As it has been explained before, there are some instructions that it is necessary to execute at each terminal opened.Then, in order to avoid future problems, it is recommendable to include the following lines in your *.bashrc*:

*setupATLAS*
*lsetup "gcc gcc493_x86_64_slc6"*
*export ILCSOFT="/path/where/you/want/to/install/EUTelescope"*
*source /afs/cern.ch/user/r/yourusername/your/path/to/v01-17-10/ILCSoft.cmake.env.sh*
*source /afs/cern.ch/user/r/yourusername/your/path/to/v01-17-10/init_ilcsoft.sh*
*source /afs/cern.ch/user/r/yourusername/your/path/to/v01-17-10/Eutelescope/v1.0/build_env.sh*
*export MARLIN_DLL=/afs/cern.ch/user/r/yourusername/your/path/to/v01-17-10/Eutelescope/v1.0/external/eudaq/v1.6.0/lib/libNativeReader.so*
*source /afs/cern.ch/sw/lcg/external/gcc/4.9/x86_64-slc6/setup.sh*

send me more information or I do

## 2.3 Recipe

**Prerequisites** :

1. ATLAS setup
2. cmake 3.3.2
3. gcc 4.9

**Summary instructions:**

1. lxplus
2. export ILCSOFT (creating a folder)
3. svn
4. Change root version at release-versions.py
5. Installing EUTelescope ( 2hours)
6. Edo branch git
7. Environment variables (three source .sh)
8. Changing EUDAQ version
9. Export to MARLIN_DLL
10. Loading all libraries (inside v1.0/build)

As an additional note, EUTelescope is a rather memory-intensive code. The initial quota in lxplus is not enough to use it for several runs. It is suggested to increase your quota on lxplus (https://resources.web.cern.ch/resources/Help/?kbid=067040) and to place the raw files in the personal workspace, where more memory is available.

## 2.4 Moving to a New Branch

As *git* works with branch, you are able to change a new branch keeping the files that you have in your branch. Currently, we have installed EUTelescope using the Edo branch, but in order to show how can you change to another branch, we have to use the following commands inside the */afs/cern.ch/user/r/yourusername/your/path/to/v01-17-10/Eutelescope/v1.0* folder. In that case we are going to change to the AlexMorton branch.

*git remote add Alex https://github.com/AlexanderMorton/eutelescope.git*
*git fetch Alex*
*git checkout -b AlexBranch Alex/addShiftToHitMaker*

So, we are at the AlexMorton branch. You can check the branch with:

*/afs/cern.ch/user/r/yourusername/your/path/to/v01-17-10/Eutelescope/v1.0$ git branch*
** AlexBranch*
*EdoBranch*
*v1.0-tag*

Now it is necessary to re-build EUTelecope. To do this, go in */afs/cern.ch/user/r/yourusername/your/path/to/v01-17-10/Eutelescope/v1.0/build* and use:

*rm -r **
*cmake ..*
*make install*

After that, we can work with this branch.

# 3 Running Some Examples

## 3.1 Basic EUTelescope files

In this section is explained which are the basic EUTelescope files needed to make it run, and how they interact with each other. Some of these will be discussed in far more detail in the following sections. The basic EUTelescope files are the following:

*.raw files*
*geometry 'gear' files (.xml)*
*.config files*
*runlist (.csv)*
*steering files (.xml)*
*source code (.cc)*

The .raw files are generated by EUDAQ and contain the event information of a single run in binary format. They are the starting point of the reconstruction and usually they should be located in the path given by the variable "NativePath" in the config file. As an additional note, the header of the .raw files contains the EUDAQ config file used in that run, and that can be check with the command: *head runXXXXXX.raw*. The .raw files will be used only during the first EUTelescope step, the converter, during which some of the information contained in them will be converted into lcio format.

The gear files contain the geometry of the set-up. In these files, it is necessary to specify the geometry of all the planes included in the conversion step. In 2016 test beams we had six Mimosa planes, one FEI-4 plane and six DUTs. The intrinsic properties of the planes, e.g. strip pitch and number of strips, have to be set in these files. If the strips are not perpendicular to the ground, it will be necessary to set a XY rotation for the DUT in the gear file. The positions and rotations will be fine-tuned during the pre-alignment and alignment in EUTelescope, so the standard consists in leaving these values as zero (apart from rotationXY, depending on the orientation of the DUT). Finally, pay attention: there is some amount of redundancy in the gear files and several parameters are repeated twice. Different processors can pick different values (but there is usually consistency within the examples). If you have a doubt regarding which parameter you should change, check other gear files to see which line had been modified there.

The user interacts mainly with the config file. Here, the paths to the other files are defined, as well as most of the variables needed by the source code. A section must be defined for every step that will be used with the command. Since this is probably the most important file, it will be discussed later together with the examples.

The steering files are a collection of .xml files, one for every step that is called with the jobsub command. Inside the steering files are defined the processors called after the jobsub command and the variables needed by those processors.

Finally, the source code can be found in */Eutelescope/v1.0/src* (and the headers in */Eutelescope/v1.0/include*). One .cc file is defined for every processor and contains the real code that runs when the processor is called. If any source file is changed, EUTelescope must be rebuilt in order to apply the changes.

The .config file, runlist and steering files interact with each other. The steering file is the one called by the jobsub command and it send directly the parameter needed by the processors.

It is possible to set these parameters in the config file. It is also possible to set parameters in the config file through the runlist. A common example is the name of the gear file for every run. An example can be seen in Figure 3.1, where the number of events to analyse in one run is defined in the runlist and sent to the source code, passing through the config and steering files.

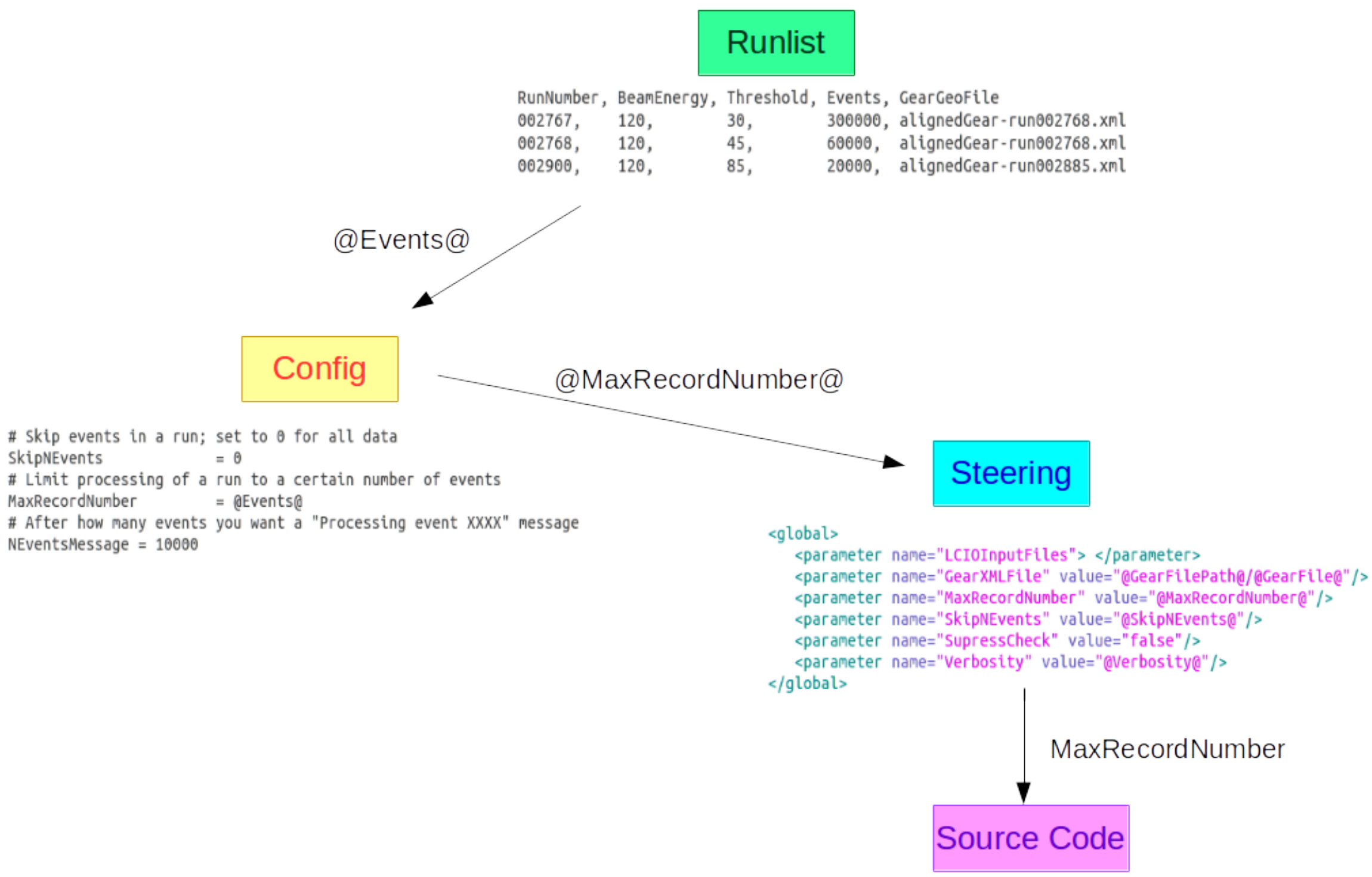

Figure 3.1: Example of a variable that is sent from the runlist to the source code, passing through the config and steering files.

In case you have problems with the steering files, you can use the command:

*Marlin -x > example.xml*

This will generate the file example.xml which contains the options of every processor known to Marlin. In this way, it is possible to check which options are available for the processor you are interested in.

## 3.2 Datura-noDUT

The most basic standard example in EUTelescope is Datura-noDUT. It is useful to run over it as a sanity check and to get a first idea on how EUTelescope works. Here, it will be explained how to run the example without explaining the basic structure of EUTelescope, that will be explained in the next sections. Moreover, this example use different processors with respect to the ones used in the 2016 ITk Strip test beams, for this reason, one should not spend too much time trying to learn and understand it.

The examples are in */Eutelescope/v1.0/jobsub/examples* and enter in *datura-noDUT*. First, it is necessary to create the folders where the output files will be saved. If EUTelescope can not find these folders, it will not be able to run:

*mkdir output*
*mkdir output/database*
*mkdir output/histograms*
*mkdir output/lcio*
*mkdir output/logs*

The folder *histograms* will contain the ROOT files with the histograms created by the various EUTelescope processors. This is the folder with the most important information and that will be used most. The folder *logs* will contain the EUTelescope output, as displayed on the terminal. The folder *lcio* will contain the collections used in (and generated by) EUTelescope in the lcio format. Finally, *database* contains lcio files with additional information that can be necessary for some EUTelescope processors, e.g. the collection of the noisy pixel.
To run a processor, the command to be used is[5]

*jobsub -c path/to/config/file -csv path/to/runlist/file processor-name run-number*

In datura-noDUT, for example, you can starting using the first processor (*converter*) for the run 97:

*jobsub -c config.cfg -csv runlist.csv converter 97*

The *converter*, converts the raw data in the lcio data format, using EUDAQ's Native Reader. Moreover, in this steps the noisy pixels are identified (but not masked). If the *converter* runs smoothly, you can then run the next processors:

*jobsub -c config.cfg -csv runlist.csv clustering 97*
*jobsub -c config.cfg -csv runlist.csv filter 97*
*jobsub -c config.cfg -csv runlist.csv hitmaker 97*
*jobsub -c config.cfg -csv runlist.csv align 97*
*jobsub -c config.cfg -csv runlist.csv fitter 97*

The functionalities of these processors will not be explained here. If one wants to have an idea of what they do, one can check the histograms in the ROOT files in the *histograms* folder.
Every processor generate an LCIO file, that will be used by the successive processor. Another way to check the functionality of every processor consist in checking the content of the LCIO file it has generated. To do this, the command *dumpevent* can be used:

*dumpevent filename.slcio event_number*

This command will show the entries of all the collections contained in the lcio file for a certain event.

[5]If we want to visualize the messages in4 our terminal with colours, we can add *-g* after the jobsub command. For instance: *jobsub **-g** -c path/to/config/file -csv path/to/runlist/file processor-name run-number*

## 3.3 ITk Strip 2016 Test Beam

## 3.4 Overview

Six steps were used in the reconstruction in 2016:

*converter*
*clustering*
*hitmaker*
*patternRecognition*
*GBLAlign*
*GBLTrackFit*

The final step will perform the tracking and produce the final .root file that will be used in the post-reconstruction analysis.
First, it is needed to align the geometry, and so obtaining a good gear file for the tracking. In order to perform the alignment, it is not necessary to use all the events in a run and a smaller number of events is usually enough. Good practice consists in using between 10000 and 50000 events. During the alignment, after having performed the first three steps, the *patternRecognition* and *GBLAlign* are iterated, until a good geometry is reached.

After this, is possible to perform the tracking with the full data set using the new aligned geometry file. When doing this, it is necessary to run again all the steps using all the events and in this case the *GBLAlign* step is not used.

When a good geometry file is generated, this can be used for all the runs within the same threshold scan, if the actual geometry was unchanged between these runs. Anyway, has been noticed that sometimes the DUT shifts by a few $\mu m$ during times of the order of hours. For this reason, it is always better to check whether there is any misalignment in the results after *GBLTrackFit*. If this is the case, a new alignment process can be needed, and so new tracking.

## 3.5 ITk Strip 2016 Example: Step by Step

Some config, gear and steering files for the 2016 reconstruction are already set in "**jobsub/example/ITkStrip2016**". This example is ready to be used out of the box. In this section some information on how to run EUTelescope is also explained.

The raw files are usually stored on the CERNbox, that uses the EOS system for storing data. It is possible to download directly the raw files from lxplus using EOS. To set up EOS on lxplus[6] is necessary to use the command:

*source /afs/cern.ch/project/eos/installation/user/etc/setup.sh*

After this is possible to explore the eos folder and copy the raw data with, for example:

*eos ls "/eos/user/a/atlitktb/CERN_July_Testbeam"*
*eos cp "/eos/user/a/atlitktb/CERN_July_Testbeam/LS3_IrradiatedLS/Data/run002768.raw" ./*
*eos cp "/eos/user/a/atlitkup/DESY_testbeam/data/run001008.raw" ./*

[6] EOS is not available on NAF with these commands

The raw files should be placed in the path defined by *NativePath* in the config file. In the example, two raw data for the reconstruction are already provided (run 2768 for LS3 and 1008 for LS4).

First of all, it is necessary to modify the runlist in order to include the new runs to be reconstructed. To do this, it is necessary to add a new row, with a value for every parameter defined in the first row of the file. Then, it is necessary to set correctly the config file, as explained in the next sections.
All the files are already set up properly in the example *ITkStrip2016* for the two example runs.

The DEFAULT section of the config file contains parameters that will be used in all the reconstruction steps. Among this, several paths must be defined, like the variable *NativePath*, that is the path to the raw files, and the *BasePath*, that directs to the main folder where the example is. The name of the gear file is the variable *GearFile* and it is defined inside the runlist. Then, *MaxRecordNumber* defines the number of events to be used during the reconstruction. For the alignment, less than 50000 events are necessary, while for the final tracking all the data set must be used. In the latter case, *MaxRecordNumber* must be set to a number higher than the total number of events in the run.
Before starting with the different processes, the flow will be the following:

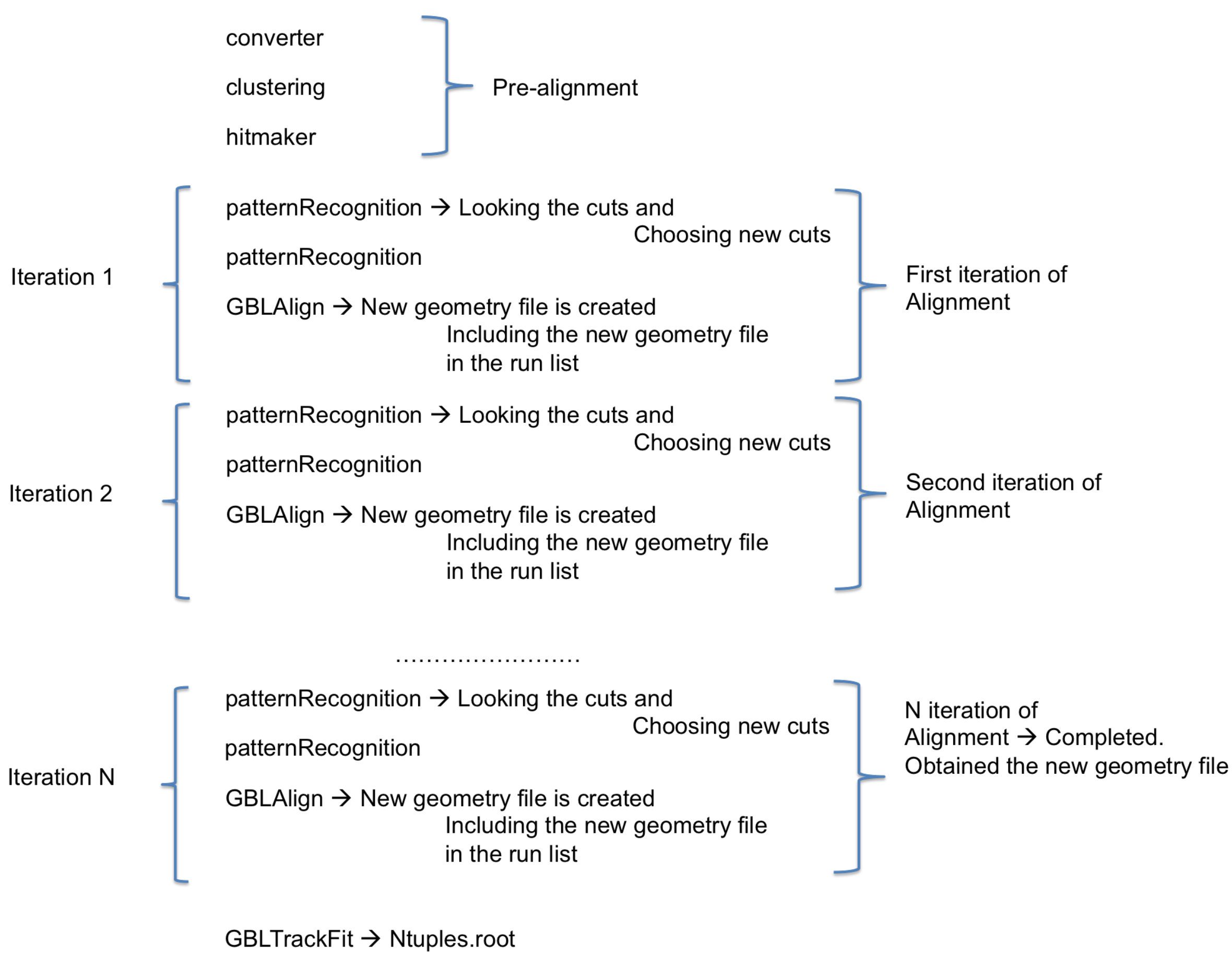

Figure 3.2: Flow of the processes until obtaining the final geometry file.

## 3.6 converter

*jobsub -c config/config_LS3_CERN_FEI4_2768.cfg -csv runlist/runlist.csv converter 2768*

The converter step decodes the raw files and converts them to the LCIO data format using the Native Reader provided by EUDAQ. Moreover, in this step the noisy pixels are identified and the collection of noisy pixels is created. This collection will be used in the clustering step to perform the masking. The only parameters that can be changed in the config file are the firing frequency cuts used to identify the noisy pixels in the MIMOSA planes, FEI4 and DUT. While for the DUT this parameter has to be left very high (for example 0.5), in order to cut only the strips with a large amount of noise, there is no real rule for the telescope and FEI4. It is costumary to leave the standard cuts in place.

$$\begin{aligned} if \quad & \frac{hits}{Number_events} > firing_frequency \\ & That\ pixel\ is\ masked \end{aligned} \tag{3.1}$$

In the Figures 3.3 and 3.4 are shown two histograms created in the converter step. In the Figure 3.3 are shown the positions of the noisy pixel in one telescope plane, while in Figure 3.4 is shown the distribution of the occupancy of the noisy pixels in that plane.

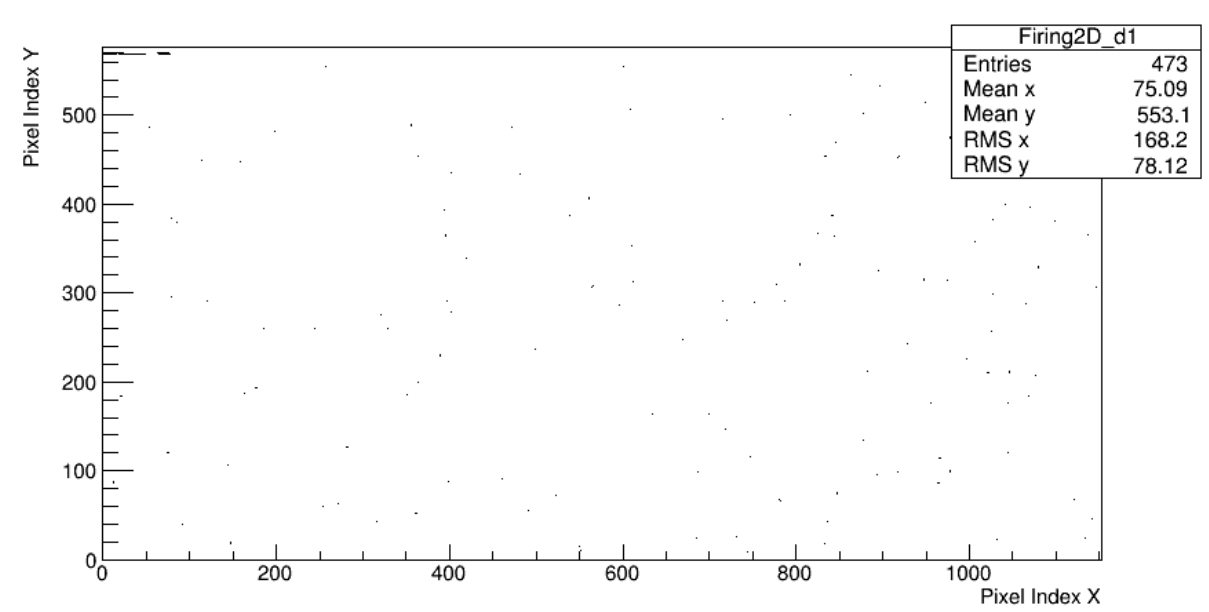

Figure 3.3: Position of the noisy pixel in the MIMOSA plane 1.

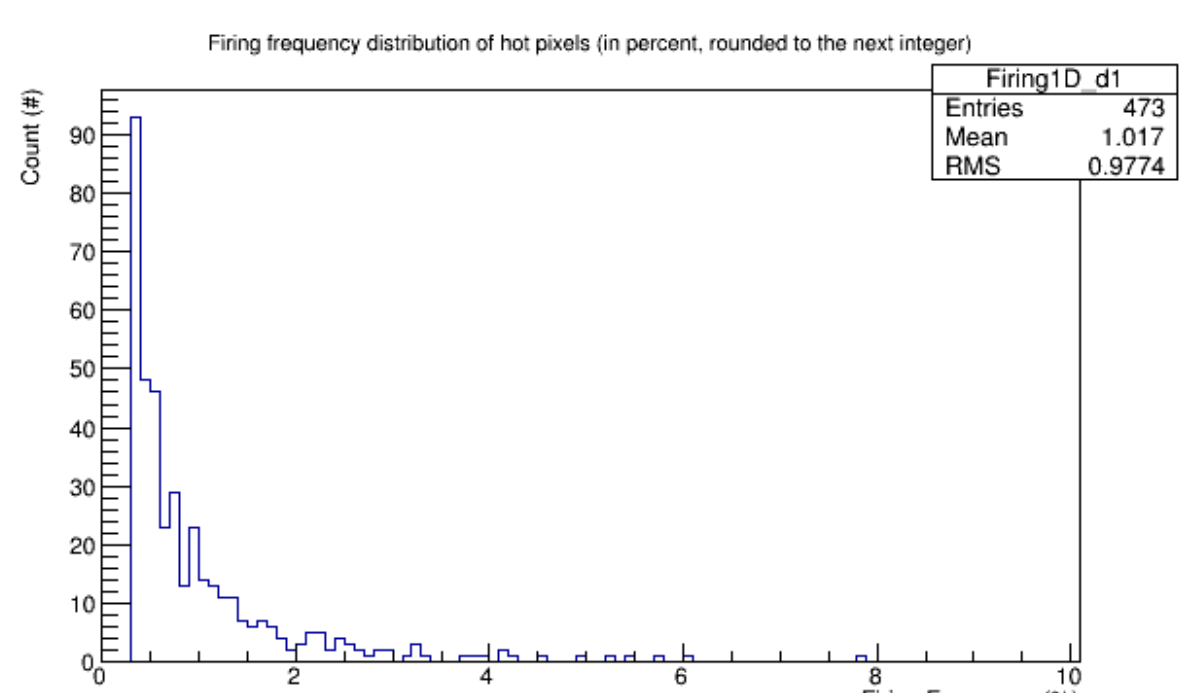

Figure 3.4: Occupancy distribution of the noisy pixels in the MIMOSA plane 1.

## 3.7 clustering

*jobsub -c config/config_LS3_CERN_FEI4_2768.cfg -csv runlist/runlist.csv clustering 2768*

In the clustering step, clusters are identified starting from adjacent hits. The noisy pixels that were identified during the converter step are masked during this step. The clustering step is rather straightforward and there are no parameters to change in the config file.

There are several interesting plots created in this step. For instance, the Figures 3.5 shows the hit map of the MIMOSA plane 0 (the axes are the pixel numbers of the hit), while 3.6 shows the cluster size distribution for the MIMOSA plane 1.

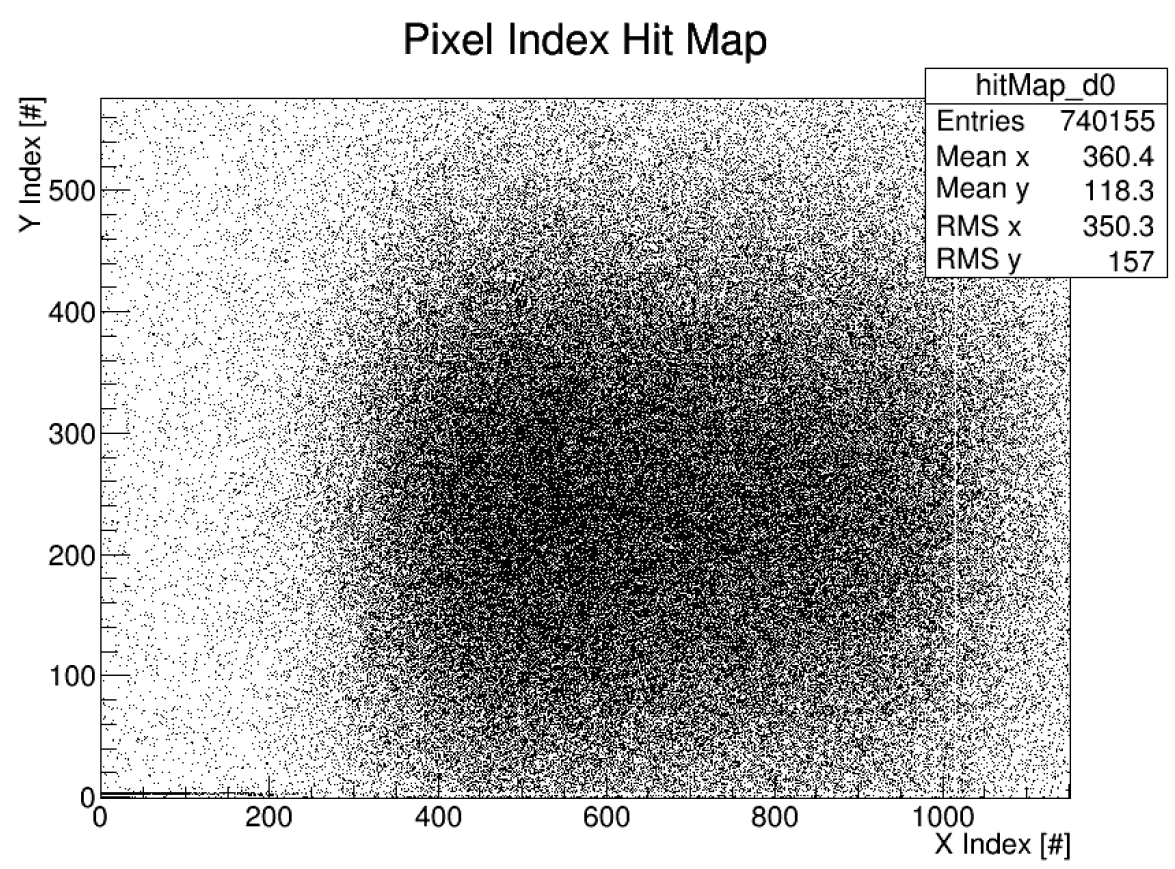

Figure 3.5: Hit map of MIMOSA plane 0.

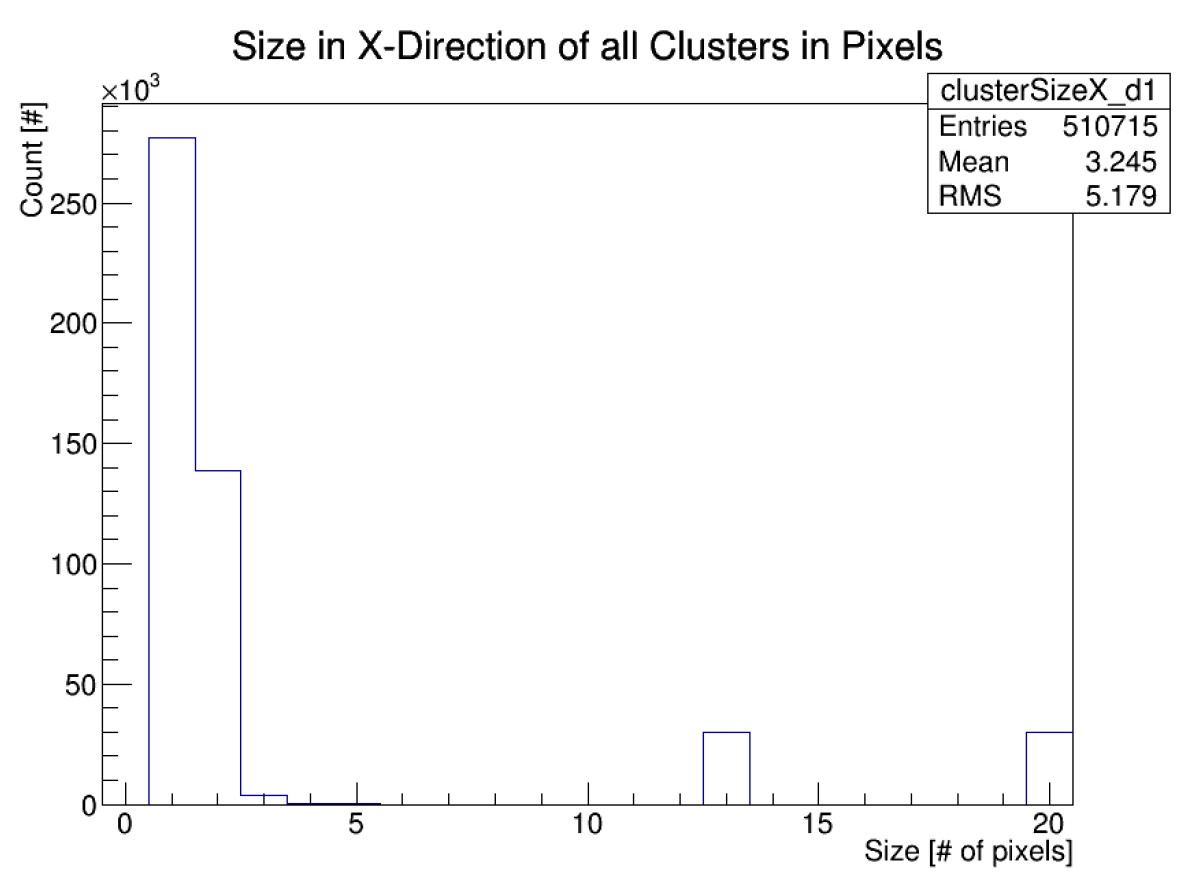

Figure 3.6: cluster size distribution for MIMOSA plane 1.

The same plots are created for the FEI4 and the DUT, as shown in Figures 3.7 and 3.8. Note that all the hit are actually in the same Y position, but they are stretched between 0.5 and 1.5 by the EUTelescope histogramming. Figure 3.8 shows the cluster size distribution of the DUT. Finally, the Figure 3.9 shows the hit map for the FEI4.

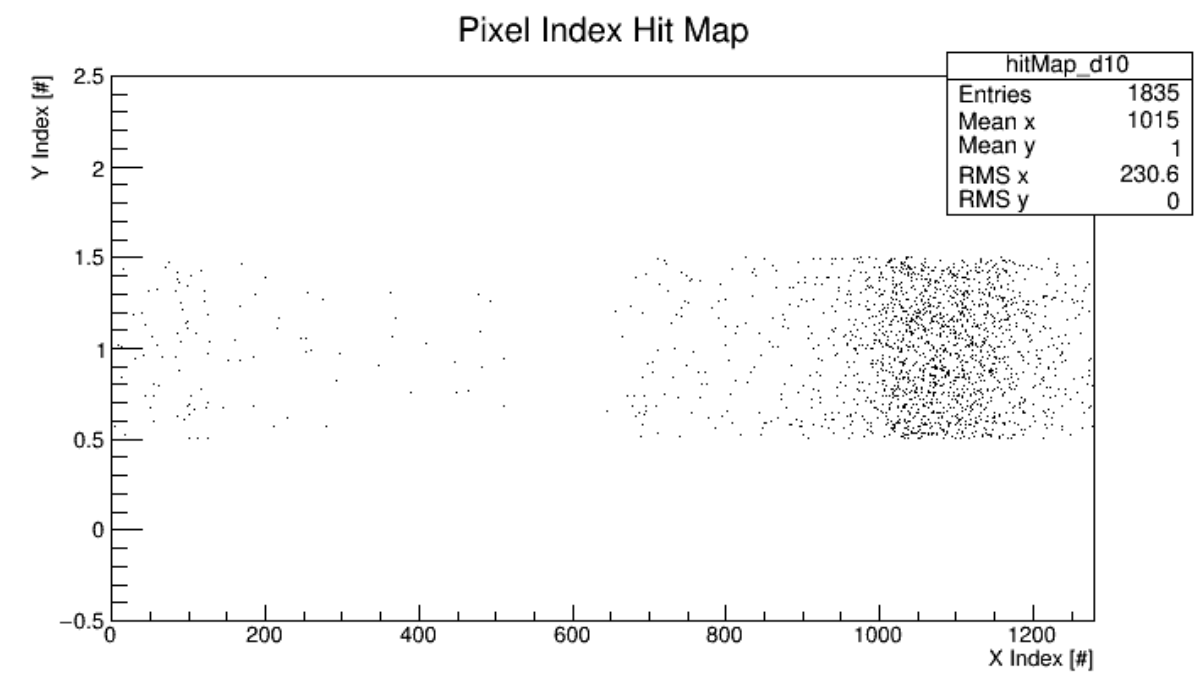

Figure 3.7: Hit map of the DUT plane 10 clusters.

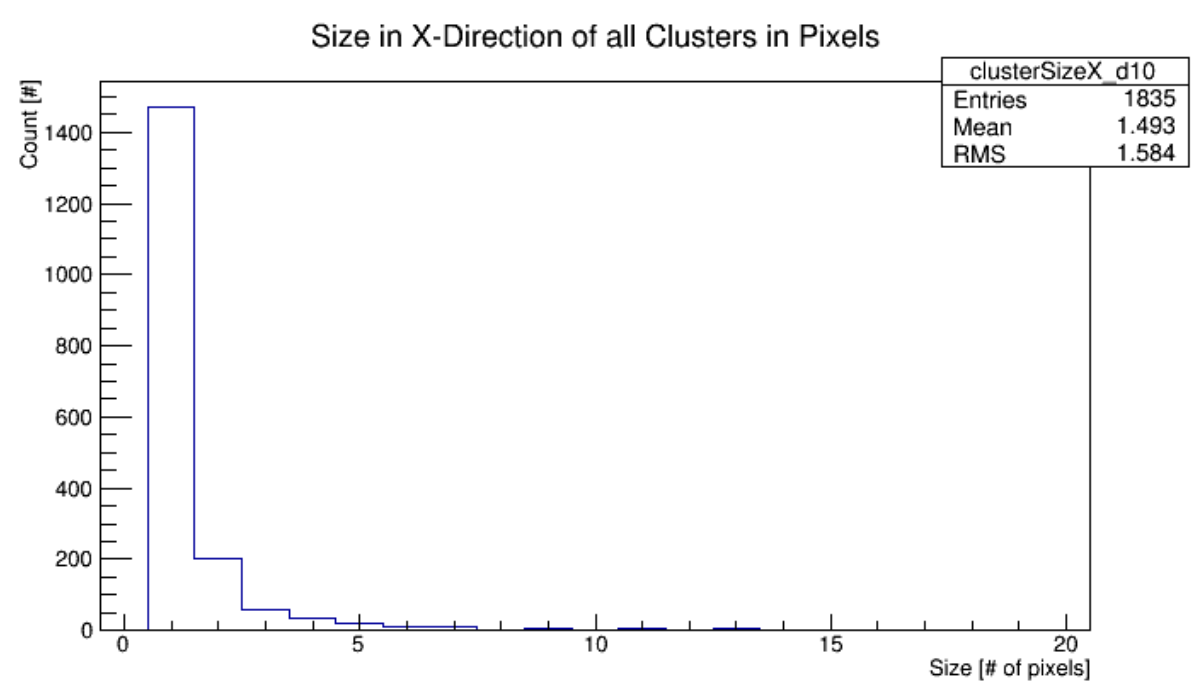

Figure 3.8: Cluster size distribution of the DUT plane 10.

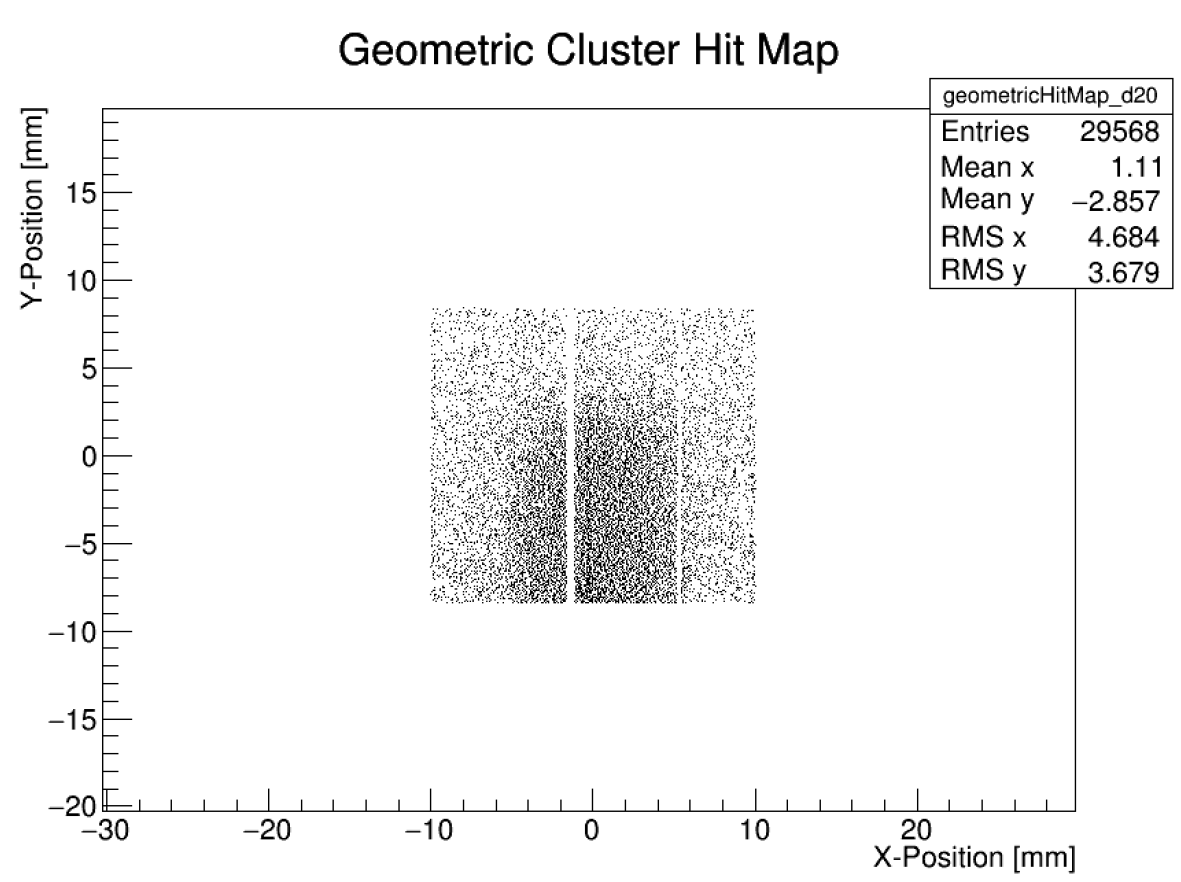

Figure 3.9: Hit map of the FEI4 plane.

## 3.8 hitmaker

*jobsub -c config/config_LS3_CERN_FEI4_2768.cfg -csv runlist/runlist.csv hitmaker 2768*

The hitmaker provides three very important functionalities:

- Creates the collection of hits in the local coordinate system.

- Correlation plots.
- Pre-alignment.

Using some of the parameters defined in the gear files, the coordinates of the hits are transformed from the pixel coordinate system to the local coordinate system (see Section 5.1 for the definitions of the coordinate systems). The collection of local hits will be used in each successive step and it has a central role in the post-reconstruction analysis.
In the config file it is also possible to change the parameter *localdistDUT*. This shifts the center of the local coordinate system from the geometrical center of the DUT. Using this, it is possible to roughly align the center of the coordinate system to the beam spot. In this way, it is possible to use tighter cuts for the prealignment.

The hits are then passed to the global coordinate system, in order to obtain the correlation plots. These are very important to check that the right rotations of DUT and FEI4 are used and as moreover these represent an important sanity check. Figures 3.10a, 3.10b and 3.10c show some correlation plots. As it can be seen in the plots, a straight line can be identified, superimposed to a more widely distributed "background". If no line is visible, it generally means that there are some problems with the geometry that is used. Often times a wrong XY rotation is the problem.

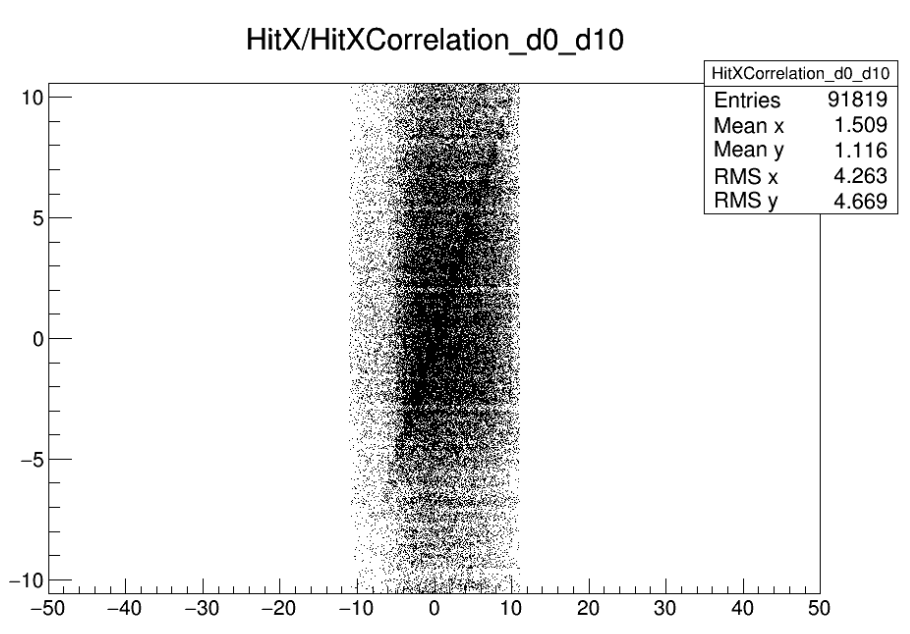

(a) Correlation between the planes 0 and 10 (DUT).

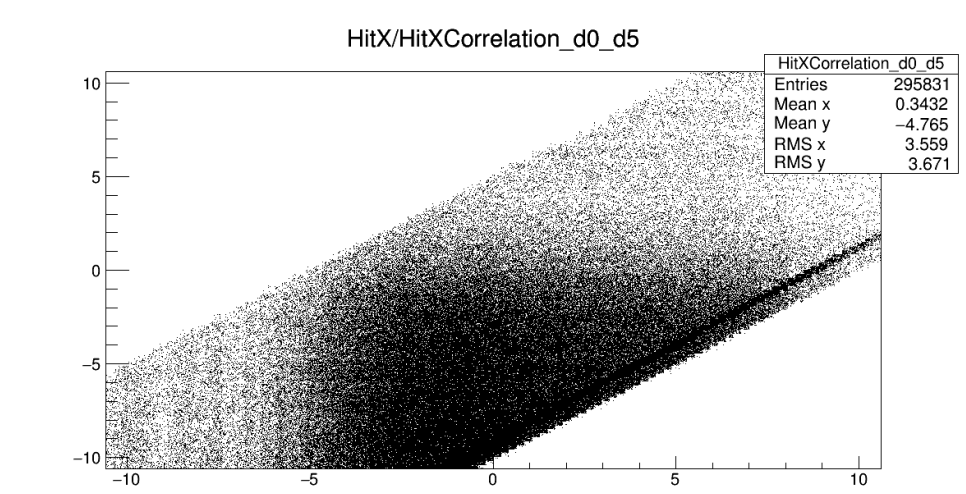

(b) Correlation between the planes 0 and 5.

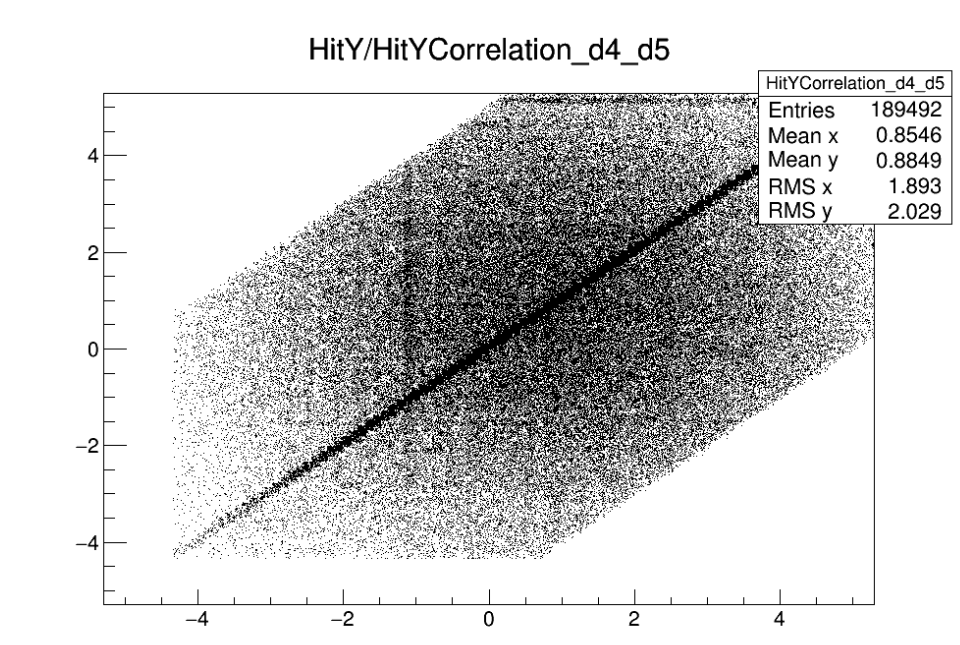

(c) Correlation between the planes 4 and 5.

Figure 3.10: Correlation between different planes.

The final functionality is the prealignment. In this step, the positions of the hits in the global coordinate system are compared between different planes. Minimizing the difference of these positions, the different planes are prealigned in the x and y directions. After this step a new gear file (called *OldGeometryFile_pre.xml*) is created with the new prealigned positions. The prealignment is very useful to have good starting estimate for the GBL alignment.

The only significant parameters in the config files are the following:

*NoEvents = 10000*
*localdistDUT = 0 0*
*ResidualsXMax = 5. 5. 5. 100. 100. 100. 100. 100. 100. 5. 10. 5. 5.*
*ResidualsXMin = -5. -5. -5. -100. -100. -100. -100. -100. -100. -5. -10. 5. 5.*
*ResidualsYMax = 5. 5. 5. 100. 100. 100. 100. 100. 100. 5. 10. 5. 5.*
*ResidualsYMin = -5. -5. -5. -100. -100. -100. -100. -100. -100. -5. -10. -5. -5.*
*ExcludedPlanesXCoord=*
*ExcludedPlanesYCoord=10*
*ExcludedPlanes = 0 11 12 13 14 15*

*NoEvents* is the number of events used both in the Correlator and in the prealignment. It is not necessary to use a full dataset for this.
*localDistDUT* is used to center the DUT with the beam spot as explained in 3.11.

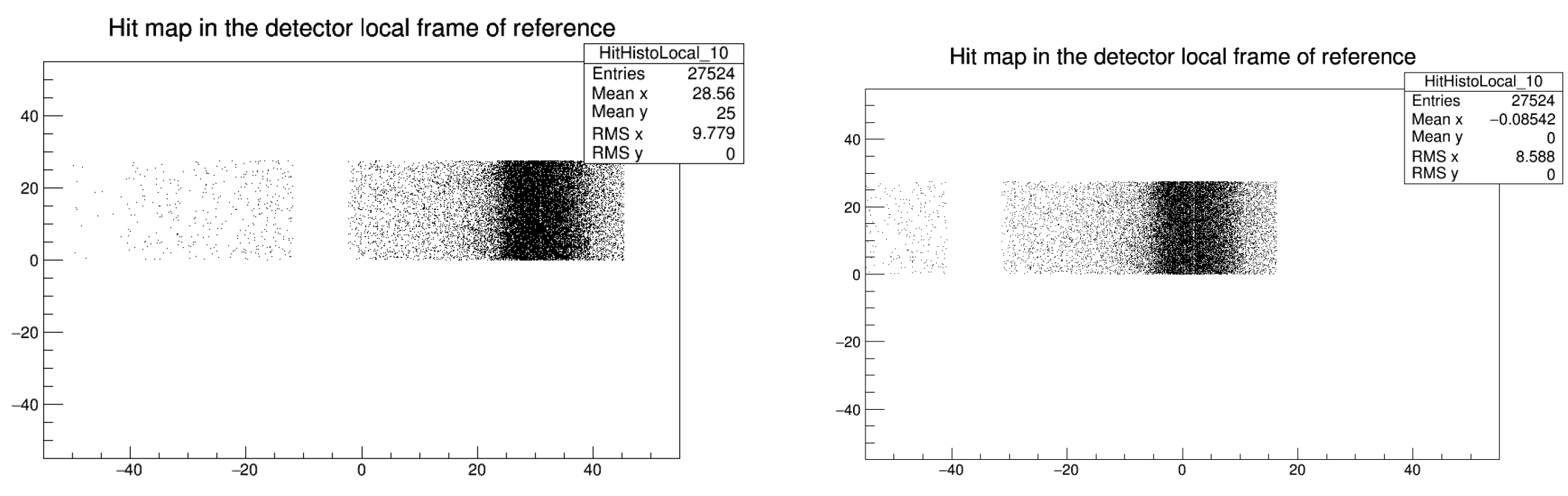

(a) Without centering (b) Centered

Figure 3.11: Hit map of the DUT without centering and centered.

The *Residual* cuts define the maximum space to look if there is a hit or not. The residuals have as reference the plane 0, so, the distances are referred to plane 0. Then, the distance between two different planes can be compared due to the reference is the same, and because of that, the correlation among two planes is possible. Furthermore, the misalignment show is also referred regarding the plane 0. The cuts are defined for every plane and the order is defined by the z positions of the planes.
The last three variables are related to planes excluded by the prealignment. It is better to exclude the coordinate that correspond to the strip direction. The excluded coordinate is in the local coordinate system.[7]

Some of the output plots are the difference in X and Y positions of the hits of one plane with respect to the hits on plane 0. The offset of the peak from 0 will be used for the prealignment. For example, Figure 3.12a show a plane with a small misalignment with respect to the plane 0, while Figure 3.12b shows a plane with a larger misalignment.

[7] The plane 0 is excluded from the prealignment, since it defines the global coordinate system.

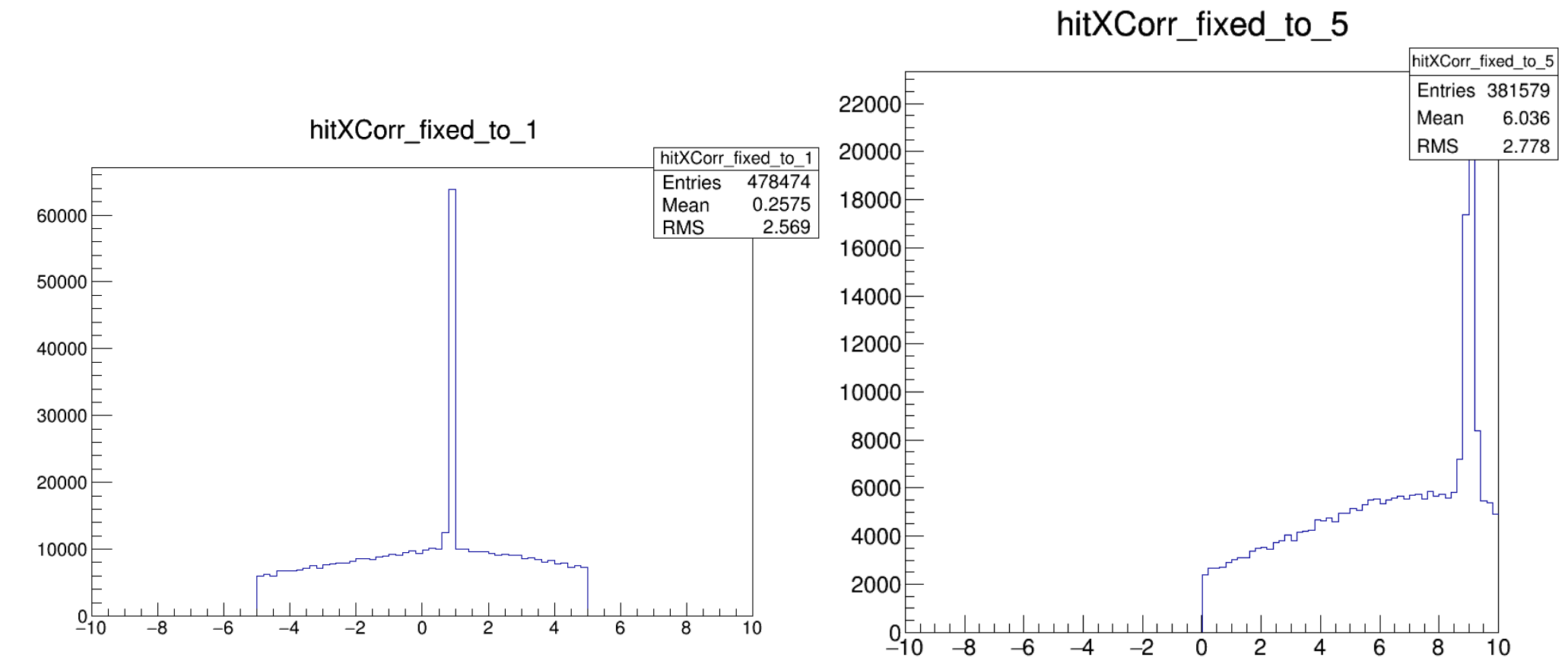

(a) Pre-alignment for the first plane. The peak is well defined with small offset.

(b) Pre-alignment for the last plane with a large offset.

Figure 3.12: Two hit correlation plots with different misalignments.

Other plots that are created are the hit maps in the local or global coordinate systems. Figure 3.13 shows the hit map for the plane 1. The beam spot is visible. An hit map of the DUT is shown in Figure 3.11. Finally, Figure 3.14a shows the hit map of the FEI4 in the local coordinates, while Figure 3.14b in the global coordinates. Comparing the two plots, it is clear that the FEI4 was rotated by 90°.

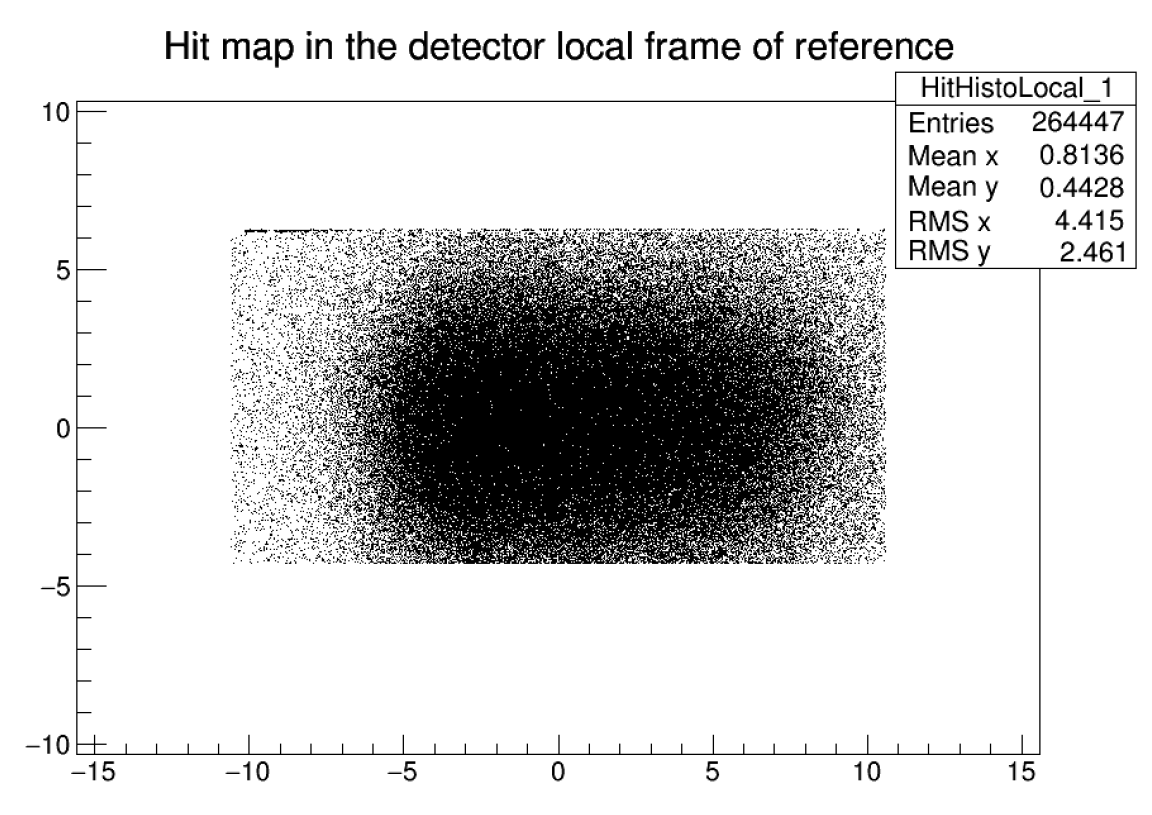

Figure 3.13: Hit map for the plane 1.

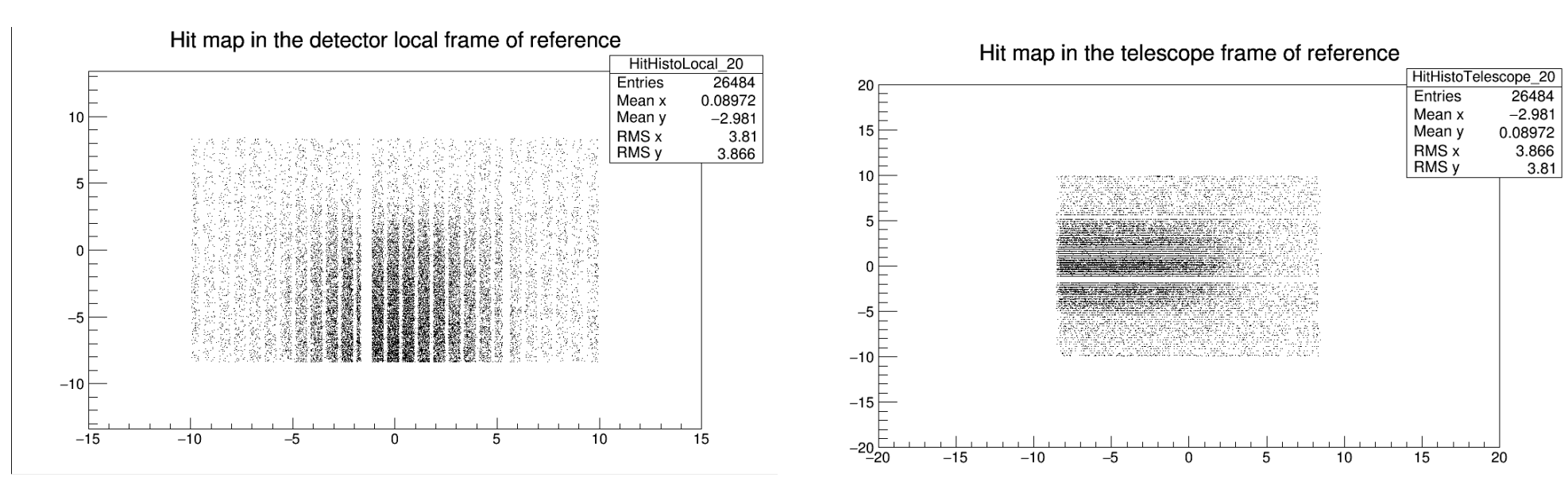

(a) Local coordinates

(b) Global coordinates

Figure 3.14: Hit map of the FEI4.

## 3.9 patternRecognition

*jobsub -c config/config_LS3_CERN_FEI4_2768.cfg -csv runlist/runlist.csv patternRecognition 2768*

The patternRecognition step groups the hits in different planes that are generated by the same track. These hits collection are then used for fitting the tracks in the GBLTrackFit and GBLAlign steps. This is probably the step where more interaction by the user is needed. It must be reminded to change the geometry file in the runlist with the new prealigned geometry files (or the geometry with the best alignment). This facilitates the pattern recognition. Notice also that the patternRecognition cuts are set in the global coordinate sytem, but the transformation from local to global coordinate system is performed independently from the same transformation in the hitmaker step.

The core of this step lies in the cuts used to identify hits that belong to the same track (and reject fake tracks). This is done used the triplets method: two different triplets are identified using the first three and the last three planes, then they are connected in the center. The cuts are the following, and they are defined in the global coordinate system:[8]

- DoubletDist: distance between the hits in the first and last plane of each three planes (planes 0-2 and planes 3-5). If some hits are within this cut, they form a doublet.

- DoubletCentDist: the hits of the doublet are connected and the position of the hit on the central planes (planes 1 and 4) is predicted. If there is a hit on that plane within this cut, then a triplet is formed.

- TripletConnectDist: using the triplets, the expected positions of the track at the center of the telescope are extrapolated. If two triplets, one upstream and the other downstream, predict a position within this cut, then the two triplets are associated with each other.

- TripletSlope: if the slopes predicted by the two triplets is within this cut, then they are considered as belonging to the same track. The collection of six hits on the six planes that has been obtained will be used in the alignment and fitting.

- DUTWindow: using the information of the other planes, the position of the track on the DUT is extrapolated. If there is a hit within this cut, than that hit will be used in the alignment step to align the DUT. This cut is used also for the FEI4 plane.

If a triplet on one arm is matched with two triplets on the other arm, then that triplet is rejected[9]. Moreover, the tracks will be saved regardless of having a hit within DUTWindow or not.
The values of the cuts are set in mm in the config file (apart from the TripletSlope cut).

[8]For further information check https://github.com/AlexanderMorton/GBLManuel/blob/master/thesis.pdf

[9]It has been noticed that decreasing the values of the cuts increases the number of tracks that passes them. The reason is that less fake tracks are taken into account, and so less triplets are rejected.

In the first iteration it is better to leave the cuts very loose (≈1 mm). In fact, every misalignment would shift the expected values of the quantities from zero. The worst the misalignment is, the larger the cuts should be. When an aligned geometry is used, the cuts should be decreased, in order to reject as many fake tracks as possible.
The value of the cuts depends on the alignment. The best way to set these cuts consists in checking the plots generated by this step, and set the cuts "by eye", as will be explained later.

Some reasonable cuts for the first and last alignment steps (the one with an aligned geometry) are the following:

$$\begin{aligned}
DoubletDistCut = 3\ 3 \quad &\Longrightarrow \quad DoubletDistCut = 3\ 3 \\
DoubletCenDistCut = 1\ 1 \quad &\Longrightarrow \quad DoubletCenDistCut = 0.03\ 0.03 \\
TripletConnectDistCut = 1\ 1 \quad &\Longrightarrow \quad TripletConnectDistCut = 0.08\ 0.08 \\
TripletSlopeCuts = 1\ 1 \quad &\Longrightarrow \quad TripletSlopeCuts = 0.005\ 0.005 \\
DUTWindow = 2 \quad &\Longrightarrow \quad DUTWindow = 0.3
\end{aligned} \tag{3.2}$$

Notice that *DoubletDistCut* has not been changed. The reason is that it can be considered a "coarse" filtering for the fake tracks, but the real "fine" selection is performed by the other cuts.
Other parameters that can be changed in the config file are the following:

*dutDirection=0*
*excludeplanes= 11 12 13 14 15*
*planeDimensions= 2 2 2 1 1 1 1 1 1 2 2 2 2*
*Planes=0 1 2 10 3 20 4 5*

*dutDirection* defines the direction where the *DUTWindow* cut should be set for mono-dimensional DUTs. 0 is used if the cut should be along X, while 1 if it should be along y. The latter one is used if the strip pitch is defined along Y instead of X.
*excludeplanes* defines the planes that are not considered in the patternRecognition, while *Planes* defines the one that should be used.
In *planeDimensions* the dimensions of the planes are set (1 for the strips, 2 for the pixels). In this way, it is decided whether to use a two-dimensional or a mono-dimensional cut for the *DUTWindow*. The order is given by the z positions of the planes, and also the excluded planes must be included in this list.

In the output, the histograms related to each cut are saved. Figure 3.15 show an example of this. Using this plots, it is possible to fine tune the cuts in order to improve the fake tracks rejection. In the following section two different ways to proceed with these cuts will be explained.

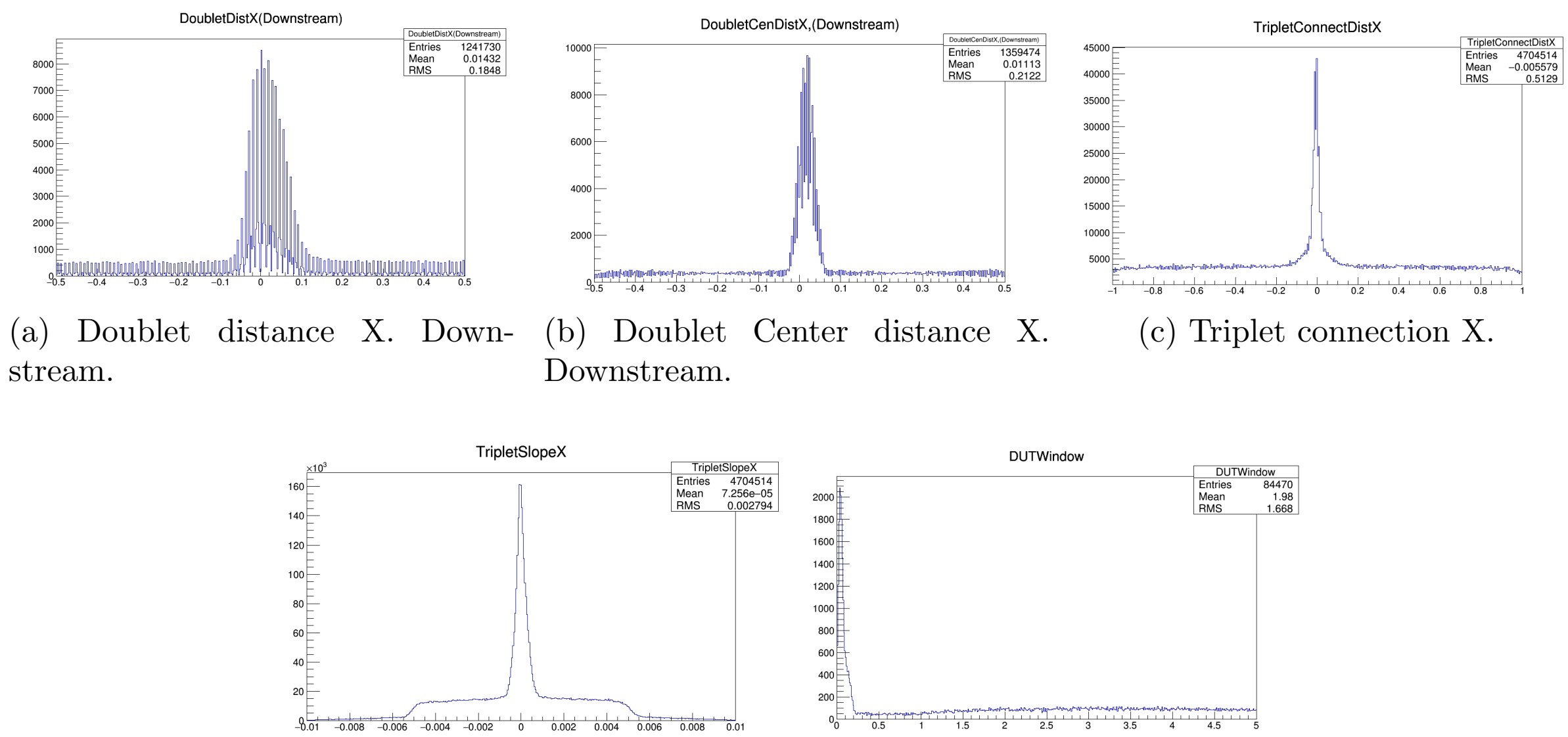

(a) Doublet distance X. Downstream.

(b) Doublet Center distance X. Downstream.

(c) Triplet connection X.

(d) Triplet Slope X

(e) DUT Window

Figure 3.15: Example of patternRecognition histograms with large cuts and some misalignment.

Other plots created are related to the number of tracks and hits on track. Figure 3.16a shows the distribution of the track multiplicity for every event. Figure 3.16b shows the hits on track candidates in every plane.

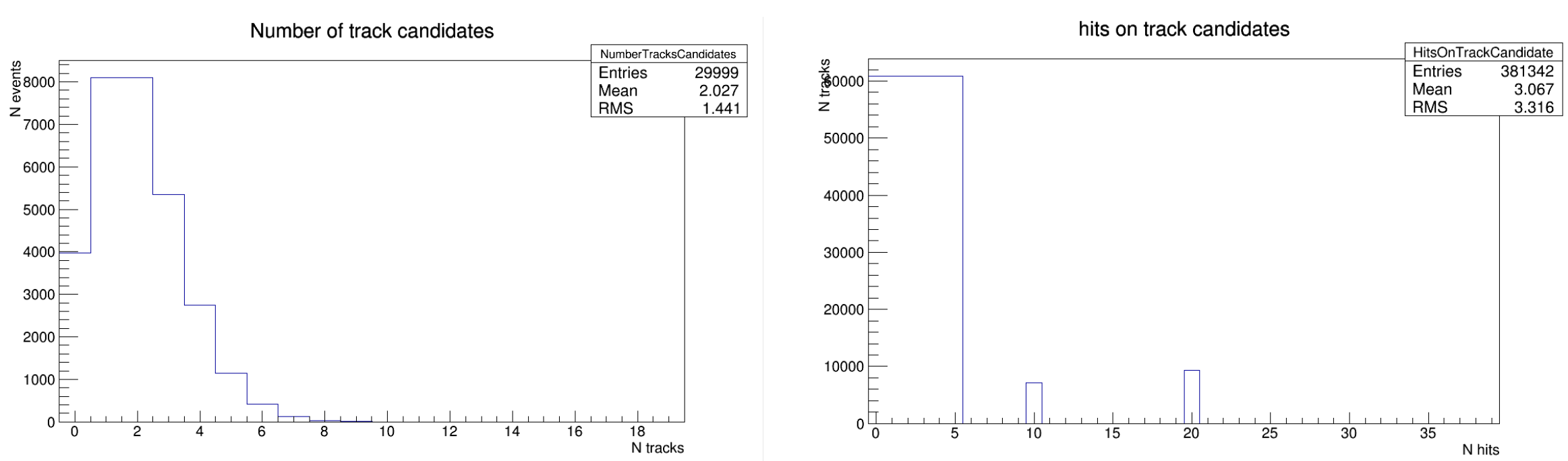

(a) Number of tracks candidates.

(b) Hits on tracks candidates in every plane.

Figure 3.16: patternRecognition plots related to the track multiplicity and hits on tracks.

As already mentioned, when a good alignment has been obtained, it is possible to decrease the cuts. In this way, more fake tracks will be rejected. Figure 3.17 shows the patternRecognition cut plots with tight cuts and a good alignment. As can be seen comparing it with Figure 3.15, the peaks look much sharper and cleaner.

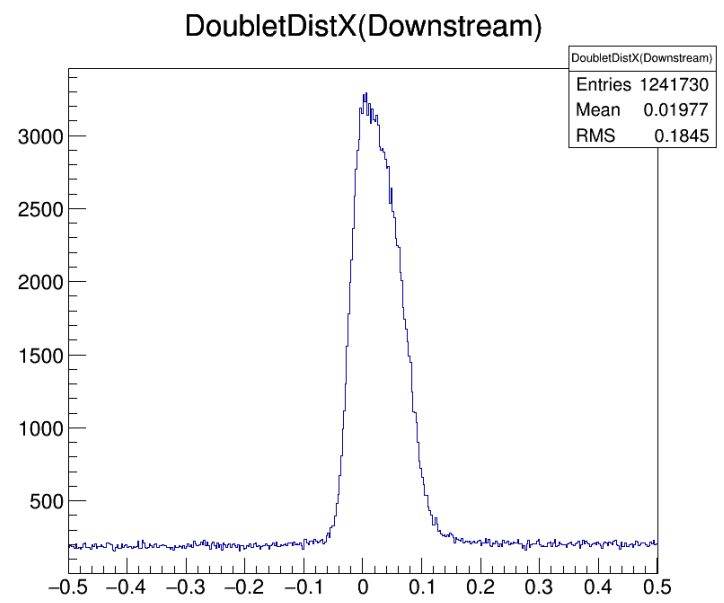

(a) Doublet distance X. Downstream.

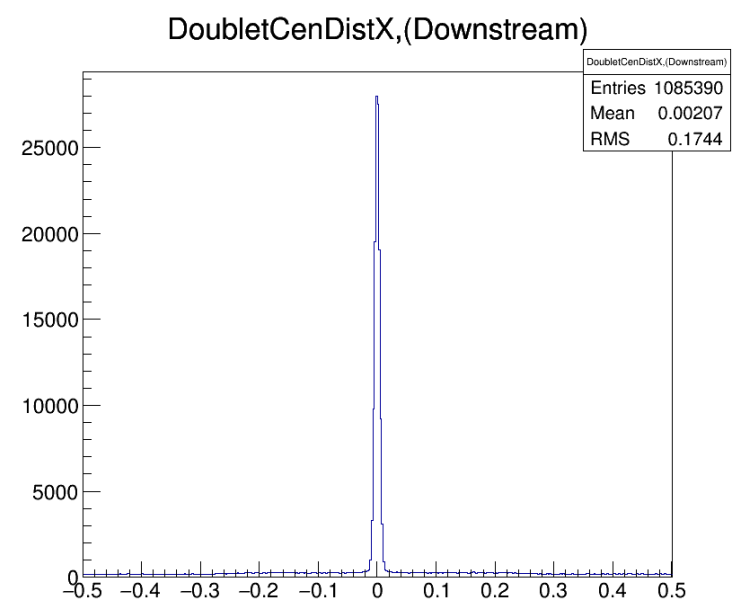

(b) Doublet Center distance X. Downstream.

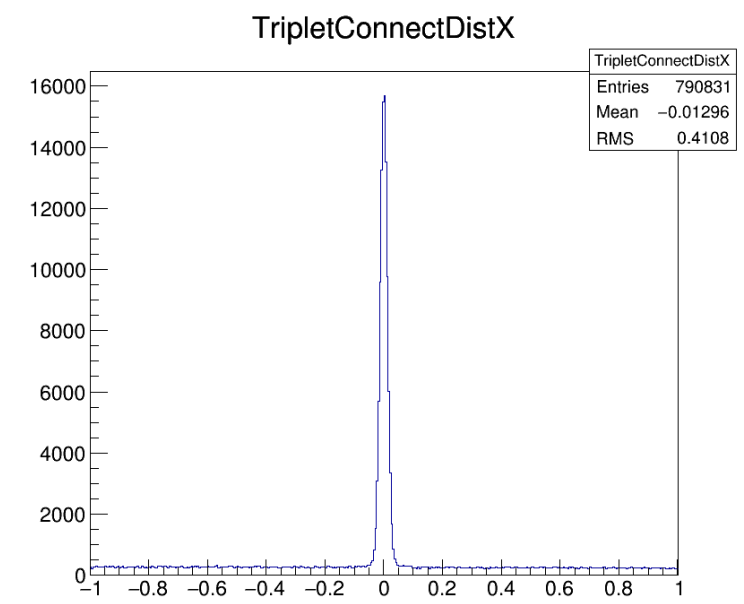

(c) Triplet connection X.

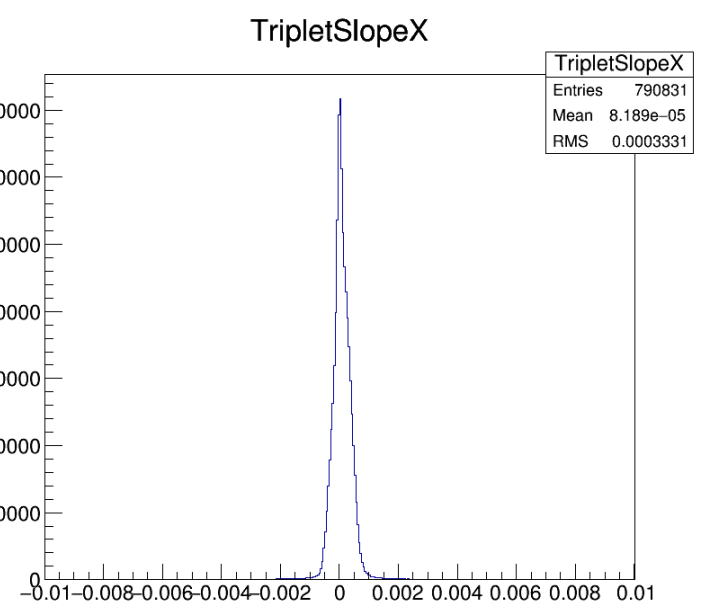

(d) Triplet Slope X

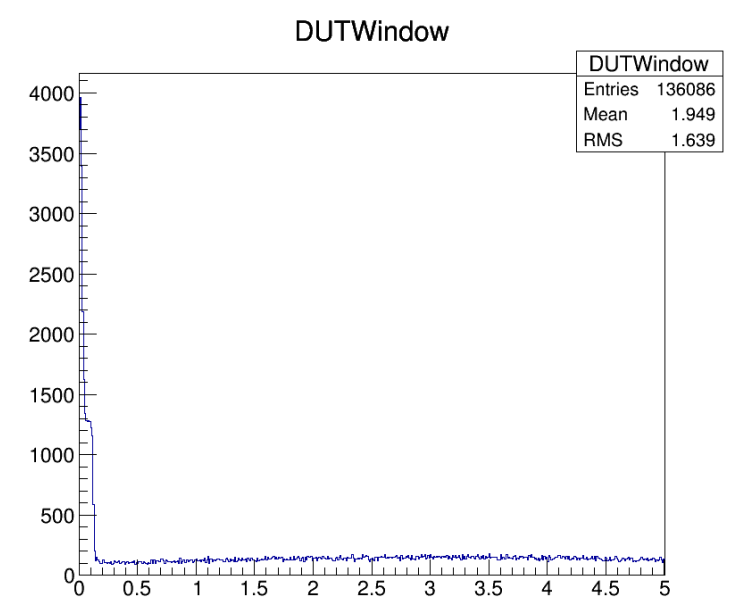

(e) DUT Window

Figure 3.17: Example of patternRecognition histograms with tight cuts and good alignment.

## 3.10 GBLAlign

*jobsub -c config/config_LS3_CERN_FEI4_2768.cfg -csv runlist/runlist.csv GBLAlign 2768*

The alignment has the role to find the right parameters for the transformation from local to global coordinate system. The groups of hits formed in the patternRecognition are fitted to obtain the tracks. Then, using these tracks it is possible to obtain the residuals, that is, the distance between the position of the track and the hit. Using these residuals it is possible to find the right alignment parameters. This is done using a chi2 minimization performed by *Millepede II*. This also sets a chi2 cut on the tracks used in the alignment.

The most important parameters that can be changed in this step are the resolutions expected for all the planes. Usually, the resolution is estimated using the standard formula:

$$\sigma = \frac{pitch}{\sqrt{12}} \tag{3.3}$$

Anyway, the clustering can improve the resolution. This effect has been measure for the MIMOSA sensors (https://arxiv.org/abs/1603.09669) and the standard values for the resolutions are the following:

*r = 0.00324*
*rFEI4X = 0.072*
*rFEI4Y = 0.0144*
*dutX=0.022*

*dutY=7.22*[10]

The resolutions are then passed to the steering file using these two parameters:

*xResolutionPlane = %(r)s %(r)s %(r)s %(dutX)s %(dutX)s %(dutX)s %(dutX)s %(dutX)s %(dutX)s %(r)s %(rFEI4X)s %(r)s %(r)s*
*yResolutionPlane = %(r)s %(r)s %(r)s %(dutY)s %(dutY)s %(dutY)s %(dutY)s %(dutY)s %(dutY)s %(r)s %(rFEI4Y)s %(r)s %(r)s*

The order is given by the Z positions of the planes. Moreover, even the planes excluded by the alignment must be inserted. Finally, X and Y correspond to the axes in the local coordinate system.

Other parameters that can be changed are the fixed coordinates:

*FixXshifts = 0 5*
*FixYshifts = 0 5*
*FixZshifts = 0 5*
*FixXrot = 0 11 5*
*FixYrot = 0 11 5*
*FixZrot = 0 5*

Usually, only the planes 0 and 5 have fixed coordinates, as requested by the definition of the global coordinate system. Sometimes, it is beneficial for the alignment to fix some coordinates for the DUT as well (as shown for *FixXrot* and *FixYrot*). In fact, since the strips are basically mono-dimensional, the rotations along the X and Y axes can not be determined very precisely. This can create problems for the alignment. If too many alignment iterations are used, one should consider to fix these rotations.

The GBLAlign step generate a new gear file in the end. The name of this file is defined in the config file:

*GearAlignedFile = name_chosen.xml*

If the name of the output aligned gear is the same as the input gear, the GBLAlign will not write save the new alignment parameters. For this reason, it is necessary to change the output name at every alignment iteration.

Finally, a list of the planes excluded by the alignment must be provided.

Figure 3.18 shows an example of the output in the terminal. The final average chi2 of the tracks is shown, as the fraction of tracks rejected because of a high chi2.

[10] The resolution in Y for the DUT is actually defined by the size of the beam. For this reason as resolution it could also be used 2.89. Anyway, this value will not change much the alignment, so it is also possible to set a very high number, in order not to consider this dimension.

```
[ MESSAGE9 "TrackAlign"]
[ MESSAGE9 "TrackAlign"]  Sum(Chi^2)/Sum(Ndf) =   265621.72340872430
[ MESSAGE9 "TrackAlign"]                       / (      556220 -          36 )
[ MESSAGE9 "TrackAlign"]                       =  0.47757886492370205
[ MESSAGE9 "TrackAlign"]
[ MESSAGE9 "TrackAlign"]             with correction for down-weighting     0.580431342
[ MESSAGE9 "TrackAlign"]
[ MESSAGE9 "TrackAlign"]
[ MESSAGE9 "TrackAlign"]       WarningWarningWarningWarningWarningWarningWarningWarningWar
[ MESSAGE9 "TrackAlign"]       arningWarningWarningWarningWarningWarningWarningWarningWarn
[ MESSAGE9 "TrackAlign"]       rningWarningWarningWarningWarningWarningWarningWarningWarni
[ MESSAGE9 "TrackAlign"]       ningWarningWarningWarningWarningWarningWarningWarningWarnin
[ MESSAGE9 "TrackAlign"]       ingWarningWarningWarningWarningWarningWarningWarningWarning
[ MESSAGE9 "TrackAlign"]       ngWarningWarningWarningWarningWarningWarningWarningWarningW
[ MESSAGE9 "TrackAlign"]       gWarningWarningWarningWarningWarningWarningWarningWarningWa
[ MESSAGE9 "TrackAlign"]
[ MESSAGE9 "TrackAlign"]         Chi^2/Ndf =   0.57999999999999996        (should be close to 1)
[ MESSAGE9 "TrackAlign"]         => multiply all input standard deviations by factor  0.76000000000000001
[ MESSAGE9 "TrackAlign"]
[ MESSAGE9 "TrackAlign"]         Fraction of rejects =  0.89000000000000001       %  (should be far below 1 %)
[ MESSAGE9 "TrackAlign"]         => please provide correct mille data
[ MESSAGE9 "TrackAlign"]
[ MESSAGE9 "TrackAlign"]       WarningWarningWarningWarningWarningWarningWarningWarningWar
[ MESSAGE9 "TrackAlign"]       arningWarningWarningWarningWarningWarningWarningWarningWarn
[ MESSAGE9 "TrackAlign"]       rningWarningWarningWarningWarningWarningWarningWarningWarni
[ MESSAGE9 "TrackAlign"]       ningWarningWarningWarningWarningWarningWarningWarningWarnin
[ MESSAGE9 "TrackAlign"]       ingWarningWarningWarningWarningWarningWarningWarningWarning
[ MESSAGE9 "TrackAlign"]       ngWarningWarningWarningWarningWarningWarningWarningWarningW
[ MESSAGE9 "TrackAlign"]       gWarningWarningWarningWarningWarningWarningWarningWarningWa
[ MESSAGE9 "TrackAlign"]
[ MESSAGE9 "TrackAlign"]
[ MESSAGE9 "TrackAlign"]  _____________________________Iteration-end________________________________
[ MESSAGE9 "TrackAlign"]    0 h  0 min 20.1 sec total                  elapsed   0 h  0 min 19.7 sec
```

Figure 3.18: Terminal output of the GBLAlign step.

Moreover, in the terminal also the changes to the alignment parameters are displayed. This is shown in Figure 3.19.

```
[ MESSAGE9 "TrackAlign"] INFO change smaller than error. ID:  1  mode:  positionX  result  -0.13674E-04  error  0.54273E-04     Removed!
[ MESSAGE9 "TrackAlign"] INFO change smaller than error. ID:  1  mode:  positionY  result  -0.28444E-04  error  0.54323E-04     Removed!
[ MESSAGE9 "TrackAlign"] INFO change smaller than error. ID:  1  mode:  rotationZY  result  0.13781E-02  error  0.19087E-02     Removed!
[ MESSAGE9 "TrackAlign"] INFO change smaller than error. ID:  1  mode:  rotationZX  result  0.99220E-03  error  0.14146E-02     Removed!
[ MESSAGE9 "TrackAlign"] INFO change smaller than error. ID:  2  mode:  positionX  result  0.60906E-04  error  0.88923E-04     Removed!
[ MESSAGE9 "TrackAlign"] INFO change smaller than error. ID:  2  mode:  positionY  result  -0.61986E-04  error  0.87744E-04     Removed!
[ MESSAGE9 "TrackAlign"] INFO change smaller than error. ID:  2  mode:  rotationZX  result  0.48651E-03  error  0.16135E-02     Removed!
[ MESSAGE9 "TrackAlign"] INFO change smaller than error. ID:  3  mode:  positionX  result  -0.41555E-04  error  0.15826E-03     Removed!
[ MESSAGE9 "TrackAlign"] INFO change smaller than error. ID:  3  mode:  positionY  result  -0.86229E-05  error  0.13919E-03     Removed!
[ MESSAGE9 "TrackAlign"] INFO change smaller than error. ID:  3  mode:  rotationZY  result  0.23065E-03  error  0.20125E-02     Removed!
[ MESSAGE9 "TrackAlign"] INFO change smaller than error. ID:  3  mode:  rotationZX  result  -0.48391E-03  error  0.18081E-02     Removed!
[ MESSAGE9 "TrackAlign"] INFO change smaller than error. ID:  3  mode:  rotationXY  result  0.57923E-06  error  0.40414E-04     Removed!
[ MESSAGE9 "TrackAlign"] INFO change smaller than error. ID:  4  mode:  positionX  result  -0.17998E-04  error  0.12150E-03     Removed!
[ MESSAGE9 "TrackAlign"] INFO change smaller than error. ID:  4  mode:  positionY  result  0.17445E-04  error  0.10825E-03     Removed!
[ MESSAGE9 "TrackAlign"] INFO change smaller than error. ID:  4  mode:  rotationZY  result  0.10830E-03  error  0.14084E-02     Removed!
[ MESSAGE9 "TrackAlign"] INFO change smaller than error. ID:  4  mode:  rotationZX  result  0.47921E-03  error  0.13535E-02     Removed!
[ MESSAGE9 "TrackAlign"] INFO change smaller than error. ID:  10  mode:  positionX  result  -0.23646E-03  error  0.63620E-03     Removed!
[ MESSAGE9 "TrackAlign"] INFO change smaller than error. ID:  10  mode:  positionY  result  0.22603E-01  error  0.34376     Removed!
[ MESSAGE9 "TrackAlign"] WARNING large rotation. ID:  10  mode:  rotationZY  result  -0.24011  error  0.20425     Removed!
[ MESSAGE9 "TrackAlign"] WARNING large rotation. ID:  10  mode:  rotationZX  result  0.43717  error  0.11995     Removed!
[ MESSAGE9 "TrackAlign"] INFO change smaller than error. ID:  10  mode:  rotationXY  result  0.59046E-04  error  0.15636E-03     Removed!
[ MESSAGE9 "TrackAlign"] INFO change smaller than error. ID:  20  mode:  positionX  result  -0.15579E-04  error  0.32598E-03     Removed!
[ MESSAGE9 "TrackAlign"] INFO change smaller than error. ID:  20  mode:  positionY  result  -0.11505E-02  error  0.12382E-02     Removed!
[ MESSAGE9 "TrackAlign"] INFO change smaller than error. ID:  20  mode:  rotationXY  result  -0.45801E-04  error  0.78610E-04     Removed
[ MESSAGE9 "TrackAlign"] The alignment parameters to update with these corrections:
[ MESSAGE9 "TrackAlign"] The sensor ID:  1  Type:  positionZ  Value:  0.21585  Error:  0.95178E-01
[ MESSAGE9 "TrackAlign"] The sensor ID:  1  Type:  rotationXY  Value:  -0.28405E-04  Error:  0.18242E-04
[ MESSAGE9 "TrackAlign"] The sensor ID:  2  Type:  positionZ  Value:  0.47153  Error:  0.16500
[ MESSAGE9 "TrackAlign"] The sensor ID:  2  Type:  rotationZY  Value:  0.27133E-02  Error:  0.20301E-02
[ MESSAGE9 "TrackAlign"] The sensor ID:  2  Type:  rotationXY  Value:  -0.87331E-04  Error:  0.31138E-04
[ MESSAGE9 "TrackAlign"] The sensor ID:  3  Type:  positionZ  Value:  0.45041  Error:  0.23363
[ MESSAGE9 "TrackAlign"] The sensor ID:  4  Type:  positionZ  Value:  0.28406  Error:  0.13837
[ MESSAGE9 "TrackAlign"] The sensor ID:  4  Type:  rotationXY  Value:  -0.31479E-04  Error:  0.23399E-04
[ MESSAGE9 "TrackAlign"] The sensor ID:  10  Type:  positionZ  Value:  1.4352  Error:  0.61340
[ MESSAGE9 "TrackAlign"] The sensor ID:  20  Type:  positionZ  Value:  0.73379  Error:  0.31285
[ MESSAGE9 "TrackAlign"] The sensor ID:  20  Type:  rotationZY  Value:  0.13901E-01  Error:  0.91086E-02
[ MESSAGE9 "TrackAlign"] The sensor ID:  20  Type:  rotationZX  Value:  -0.40250E-02  Error:  0.22548E-02
```

Figure 3.19: Changes to the alignment parameter shown in the terminal.

No output histograms will be saved by the GBLAlign step.

### 3.10.1 Iterations between patternRecognition and GBLAlign

To find an aligned geometry, several iterations of the patternRecognition and the GBLAlign are needed. Each iteration will have better starting guesses for the alignment and a lower amount of fake tracks, because of tighter cuts in the patternRecognition. The number of iterations needed depends strongly on the initial geometry used, prealignment, patternRecognition cuts, beam and noise.

There are two different ways to proceed:

- Progressive decrease of resolution
- Double patternRecognition step

The main problem lies in the fact that Millepede II does not work if the fraction of rejected tracks is over 30%, as shown in Figure 3.20. This would mean that more than 30% of the tracks have a chi2 of more than 6 (the minimum chi2 cut used by Millepede II as a standard).

```
[ MESSAGE9 "TrackAlign"]
[ MESSAGE9 "TrackAlign"]  Data rejected in previous loop:
[ MESSAGE9 "TrackAlign"]               0  (rank deficit/NaN)          0  (Ndf=0)        21299  (huge)            0  (large)
[ MESSAGE9 "TrackAlign"]  Too many rejects (>33.3%) - stop
[ MESSAGE5 "TrackAlign"]
[ MESSAGE5 "TrackAlign"] Number of rejects high. We can't use this binary for alignment
[ MESSAGE9 "TrackAlign"] THE NUMBER OF REJECTED TRACKS IS TOO LARGE.
```

Figure 3.20: Error due to a too large number of rejected tracks.

This is a problem for the first alignment iterations, where the patternRecognition cuts are very loose, and so many fake tracks are accepted, and there is some misalignment that increase the average residuals.

With the first approach, loose cuts are set in the first patternRecognition step, and a very large telescope resolution is set in the alignment step. In this way, the chi2 calculated for the tracks is reduced. This will prevent Millepede II from rejecting too many tracks, but more fake tracks will be included. At this point, at every iteration both the cuts of the patternRecognition and the telescope resolution are decreased, until the right resolution are reached. To change the cuts in the patternRecognition, the plots provided by the previous iteration are used.

In the second approach, a first patternRecognition step is followed by a second patternRecognition step with cuts already fine-tuned, based on the plots provided by the first one. In this way, the number of fake tracks is heavily reduced. Still, it is possible that is necessary to increase the telescope resolution during the first iterations. In the second iteration, again two patternRecognition steps are used, with the second one having cuts based on the plots of the first one. At this point the resolutions to be used in the GBLAlign are already the right ones. With this approach many more patternRecognition steps are needed, but it has been showed that it is possible to reach a very good alignment already in two iterations. The flow to proceed in this way is shown in the figure 3.2.

## 3.11 GBLTrackFit

*jobsub -c config/config_LS3_CERN_FEI4_2768.cfg -csv runlist/runlist.csv GBLTrackFit 2768*

After a good alignment has been found, it is necessary to perform the entire reconstruction chain using the full dataset and the new aligned geometry. It means that in that case, the parameter *MaxRecordNumber* should have a large value, 1000000 for instance, in order to include all events. All the steps must be run[11] except GBLAlign: GBLTrackFit is performed after the patternRecognition. This step reconstruct the tracks and generate the final .root file

[11]Note that although hitmaker creates a new geometry file, in this case the geometry file set in the runlist must not be changed.

that is used for the post-reconstruction analysis.

As in GBLAlign, the resolutions of the planes must be defined in the config file. These are used to evaluate the chi2 of the tracks. Moreover, the name of the final .root file must be defined:

*OutputPathROOT=NameOfTheROOTFile.root*

This will be saved in the folder where the jobsub command has been called.

There are several interesting plots generated by GBLTrackFit. The most important are the residual plots, as shown in Figure 3.21 and 3.22. The residuals are defined as the difference between the position of the track, as expected by the fit, and the position of the hits.

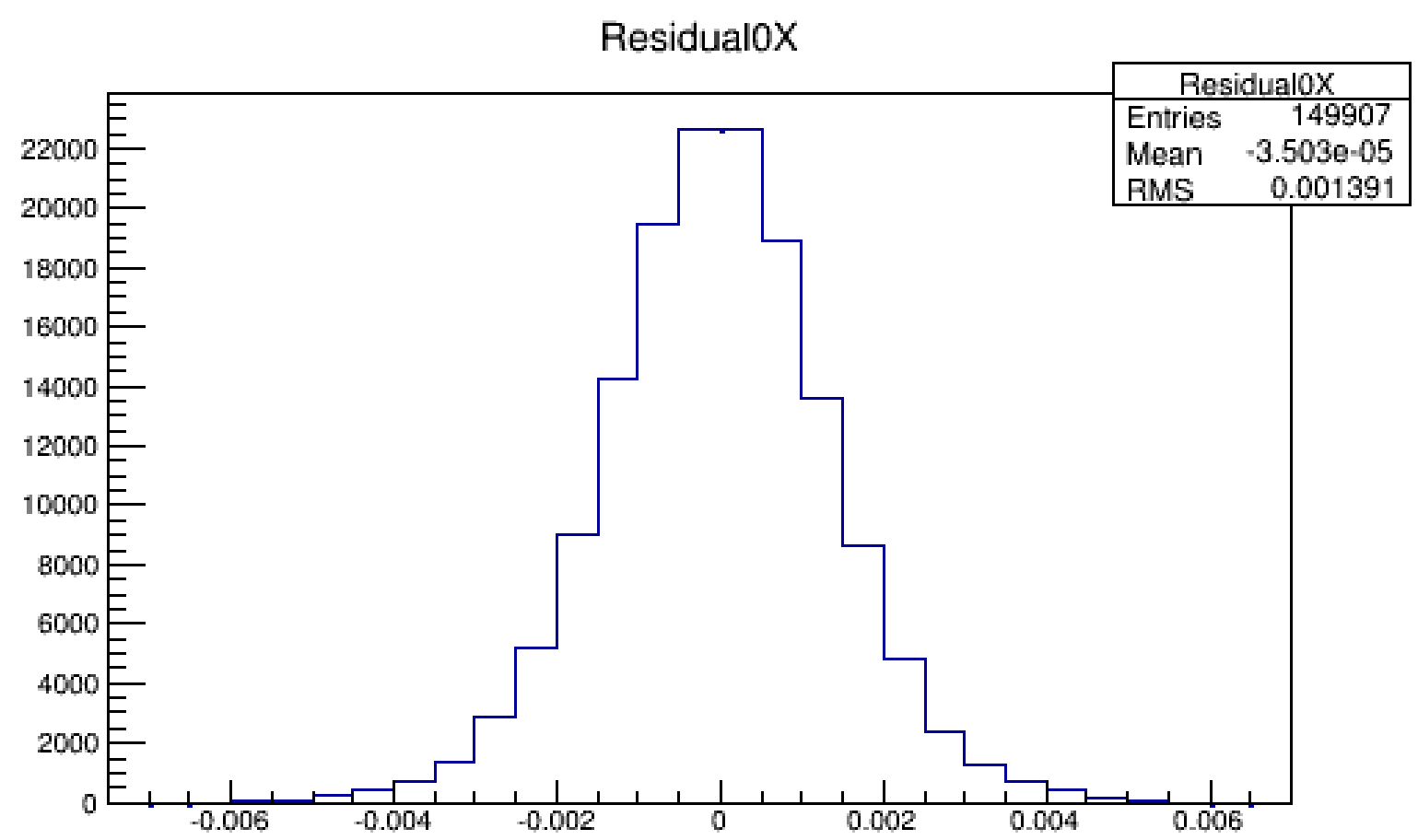

Figure 3.21: Residuals of the MIMOSA plane 0 along x.

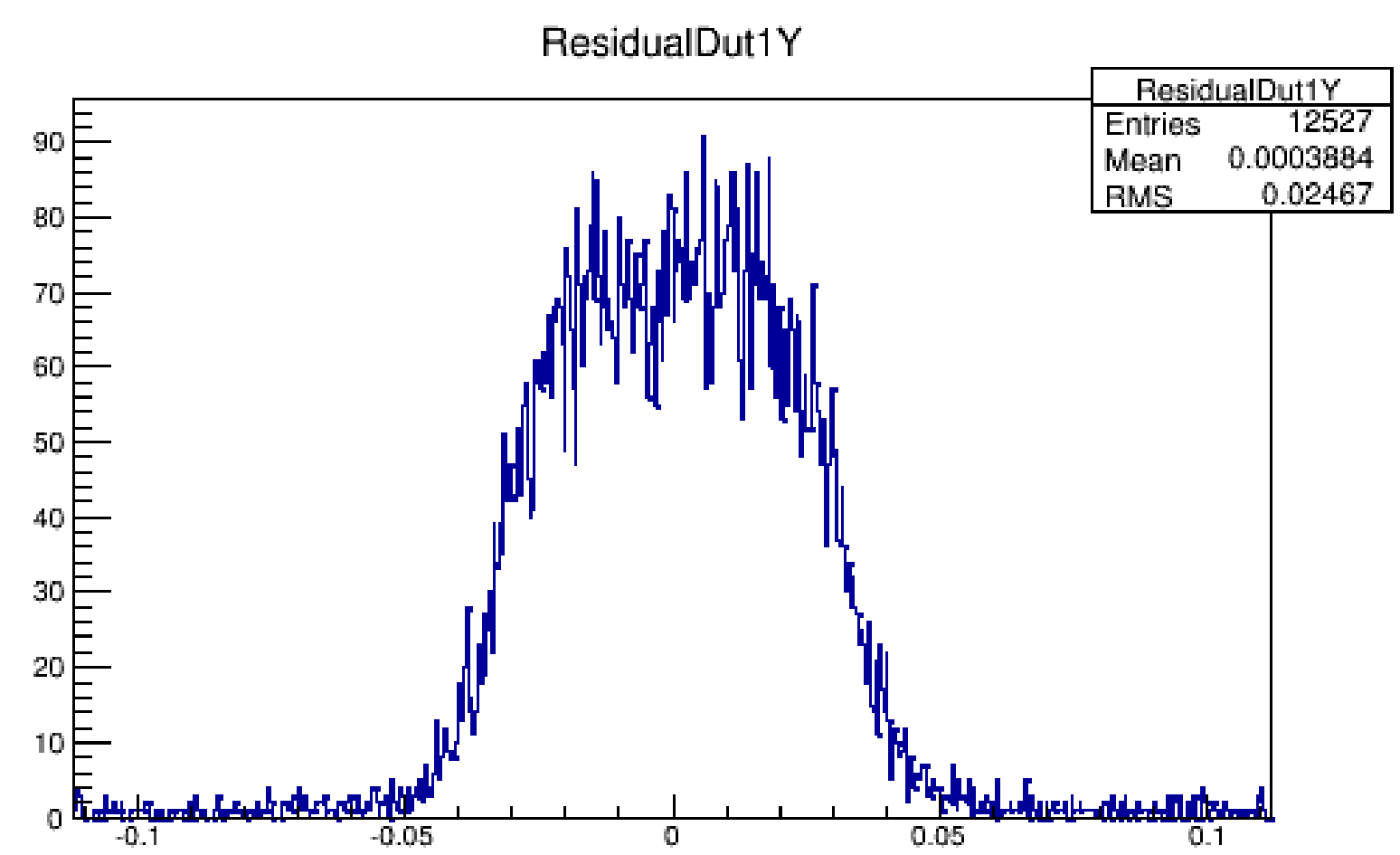

Figure 3.22: Residuals for the DUT along y (the orientation of the strip pitch for this specifical run).

The mean of the residuals should be very close to 0 if the alignment is correct. Any large deviation from 0 correspond to a misalignment. As a rule of thumb, a good alignment produces average residuals on the DUT of less than 2 $\mu$m. The standard deviation of the residuals it is related to the intrinsic resolution of the planes and to the resolution of the tracking.

If the hit information of a plane is included in the tracking, the expected standard deviation is the one for biased residuals:

$$\sigma^2_{residuals} = \sigma^2_{intrinsic} - \sigma^2_{tracking} \tag{3.4}$$

This is the case of the MIMOSA planes, that are used for the tracking. If the hit information of the plane is not included in the tracking, the standard deviation is the one for unbiased residuals:

$$\sigma^2_{residuals} = \sigma^2_{intrinsic} + \sigma^2_{tracking} \tag{3.5}$$

This is the case of the DUT and FEI4[12]. Using this formula, it is possible to estimate the resolution of the tracking.

It is worth to mention that to obtain both the mean and the standard deviation of the residuals often it is necessary to perform a gaussian fit on their distribution. In this way the contribution of the long tails produced by noise hits is minimized.

Other interesting plots that are produced in the GBLTrackFit are the hit maps of the tracks (Figure 3.23) and their chi2 distribution (Figure 3.24).

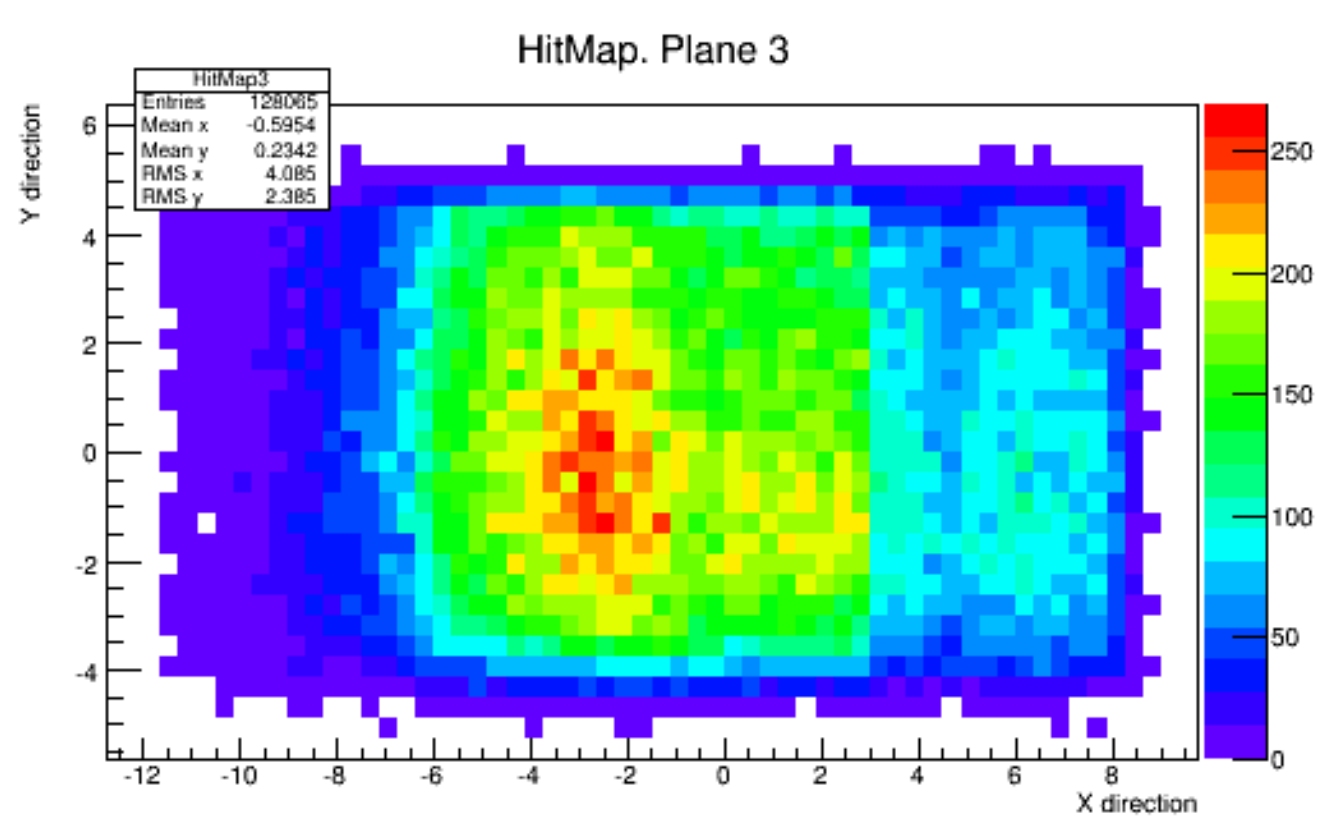

Figure 3.23: Hit map of the tracks on the plane number 3.

[12]The DUT hit information is not used in the tracking in order not to bias the track positions. The tracks are obtained only using the MIMOSA hits, and in this way their position on the DUT is extrapolated.

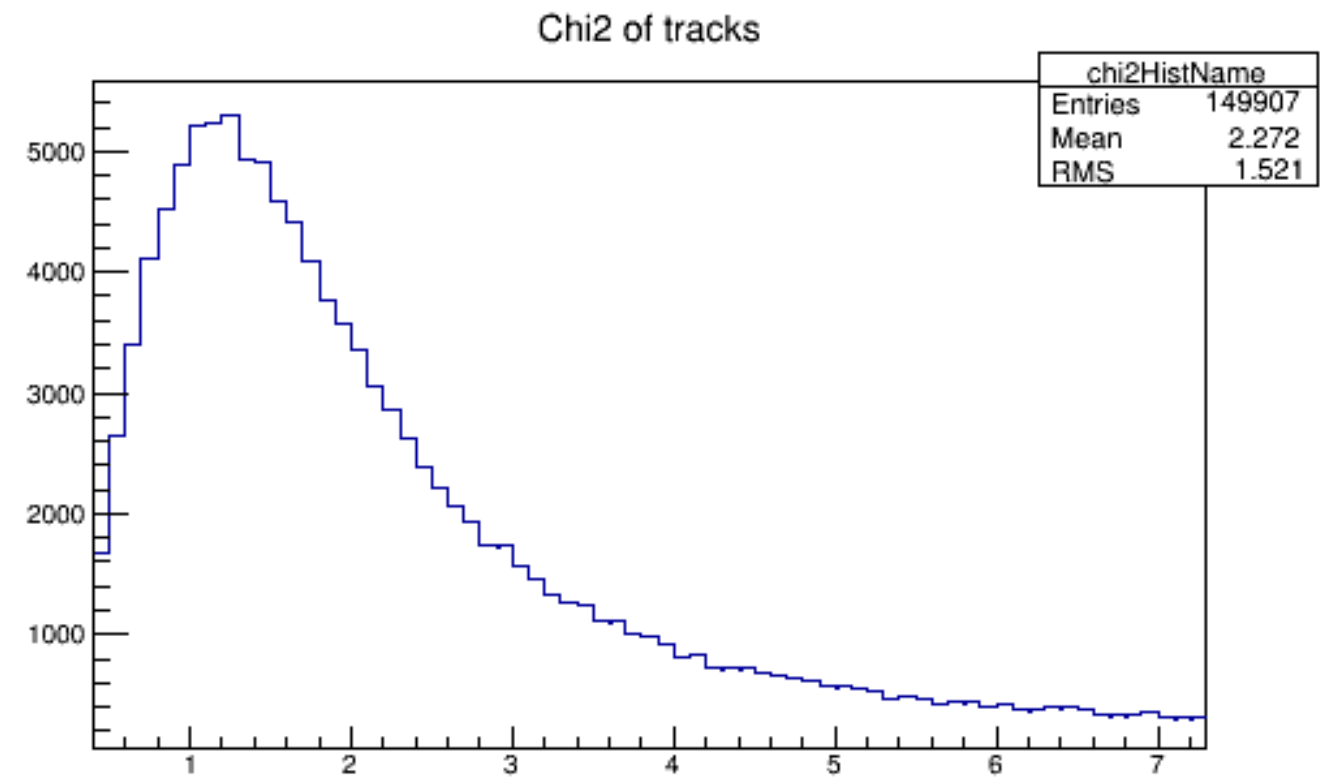

Figure 3.24: Chi2 distribution of the tracks.

As it can be seen, the mean of the chi2 distribution obtained in 2016 differs from the expected value of 1.

# 4 Post-Reconstruction Analysis

The GBLTrackFit step produces a .root file that contains all the significant information regarding the hits and the tracks. Analyzing this file, it is possible to obtain the desired results. The analysis code used in 2016 can be found at `https://gitlab.cern.ch/miqueits/TestbeamAnalysis`.

The TTrees used in the analysis are the following:

- zsdata_m26, zsdata_apix, zsdata_strip: all the strips and pixels that have fired, in the pixel coordinate system.
- local_hit: the hits of all the planes in the local coordinate system.
- GBL_tracks: the extrapolated position of the tracks in the local coordinate system.
- zsdata_strip_TTC: several timing information of the DUT, in particular the TDC time needed for applying the timing window.

The analysis code is divided in two parts. The first parts analyses single runs and it filters the data, creating a new .root file with a new TTree that contains only the information needed for the analysis. Moreover, this code creates several histograms that can be used for a sanity check or for having fast results. The second code several .root files created by the first one to create the s-curves. In this step, the calibration of the ASIC is used.

To run the code the *setupATLAS* is needed, as the right ROOT configuration. To obtain them on lxplus use the commands:

*setupATLAS*
*localSetupROOT*

With ROOT 6 it will be impossible to perform the reconstruction with EUTelescope. For this reason, if the command *localSetupROOT* has been used, it will be necessary to return to ROOT 5 to reconstruct other data.

## 4.1 First Analysis Code: Filtering

To run the first code, it is necessary to setup ROOT 6[13] and then running the command:

*./make_testbeam_analysis.sh --config settings.config --input ntuple.root*

The .root file is the final output of EUTelescope, while the config file is the one that contains the parameters that can be changed in the analysis (it should not be confused with EUTelescope config file). Figure 4.1 shows an example of such config file.

[13] This version is already loaded with localSetupROOT.

```
DUT_ID:  13
FEI4_ID: 20
DUT_FEI4_pos: 3

# chi2 requirement on GBL tracks
GBL_chi2: 10

# Timing window
min_time: 0
max_time: 50

# Strip pitch and beam offset (in mm)
strip_pitch: 0.0745
beam_offset: 80

pix_cut_min: 1036
pix_cut_max: 1120
y_cut_min: -4.6
y_cut_max: 0

# DUT rotated by 90 degrees relative to FEI4
rotated: 1

# Noisy/inefficient pixels to be vetoed
#bad_pixels:  100 200 300
```

Figure 4.1: A config file used in the post-reconstruction analysis.

Some of the parameters are trivial to understand, and they will not be explained here.
*DUT_FEI4_pos* is the number of planes between the FEI4 and the DUT plus one. Only the planes that were used in GBLTrackFit are counted. For example, in the run to which Figure 4.1 is referred, the DUT was placed after the plane 2, while the FEI4 after the plane 4. In this case *DUT_FEI4_pos* is then 3.
*min_time* and *max_time* define the timing window used in this code. This is used just by the plots created by this code, but the cut is not applied to the data stored in the .root file. For this reason, it is necessary to apply a second independent time window in the second code.
*beam_offset* is related the parameter *localdistDUT* set in the hitmaker section of the EUTelescope config file. It can be calculated with the formula:

$$beam_offset = \frac{sensor_size}{2} - localdistDUT \tag{4.1}$$

This is used to transform the hit information from the local coordinate system to the global coordinate system. *sensor_size* corresponds to the size set in the gear file and this doesn't necessarily correspond to the physical or active dimensions of the sensor.
*pix_cut_min* and *pix_cut_max* correspond to the strip numbers of the cuts that are going to be set along the direction of the strip pitch. *y_cut_min* and *y_cut_max* are the cuts in the direction of the strip and their values are in mm. If these cuts are needed, it is suggested to run a first time the code, then run it again applying the cuts based on the plots produced by the first iteration.
*rotated* must assume a value of 0 if in the gear file the strip pitch has been defined along the X direction. It should assume a value of 1 if they were defined in the Y direction.
*bad_pixels* is a list of strips that will not be considered in the analysis. These strips can be noisy or inefficient, for example. The code will reject the tracks that pass on these strips and on their neighbors.

After running the first code, a new .root file will be created. It contains a TTree that can be analysed with the second code or with an independent code. Moreover, it also contains several interesting histograms. For example, the profile of the efficiency along the X and Y axis are shown in Figure 4.2.

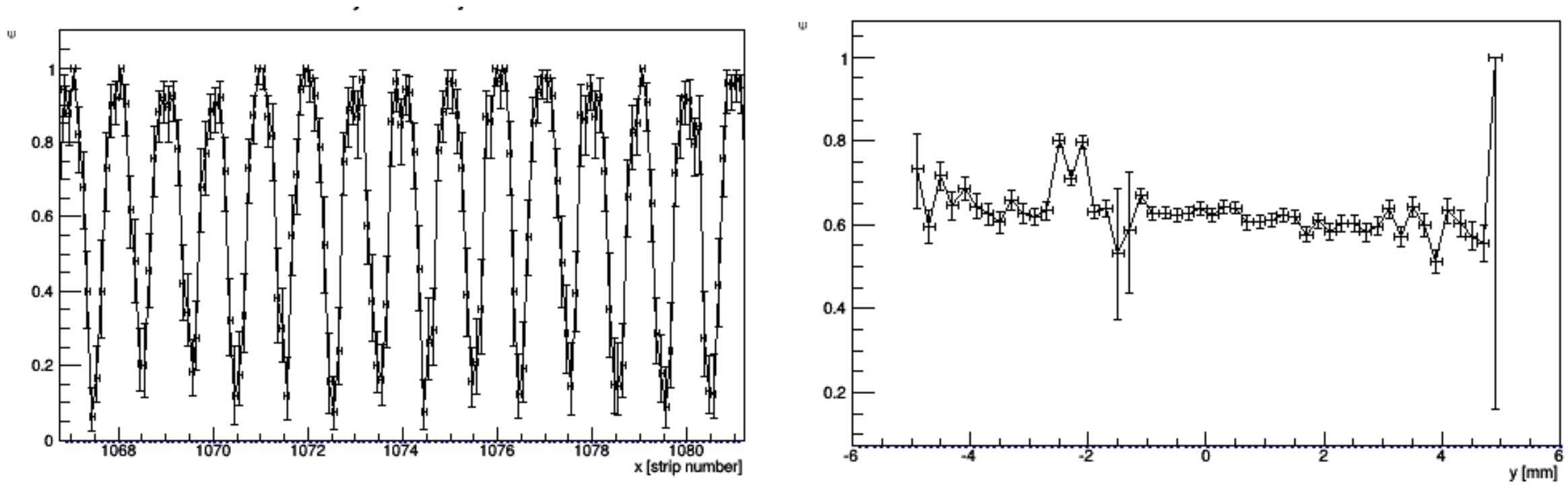

(a) Zoom of the efficiency profile in the x direction. (b) Efficiency profile in the y direction.

Figure 4.2: Efficiency profiles at a low threshold.

The X axis is always in the direction of the strip pitch in these plots, independently from the definition in the gear file. Other plots are related to the probability of having a cluster with size greater than 1[14]. An example of such a plot for a run with low threshold is shown in Figure 4.3, where bumps in the interstrip region are clearly visible.

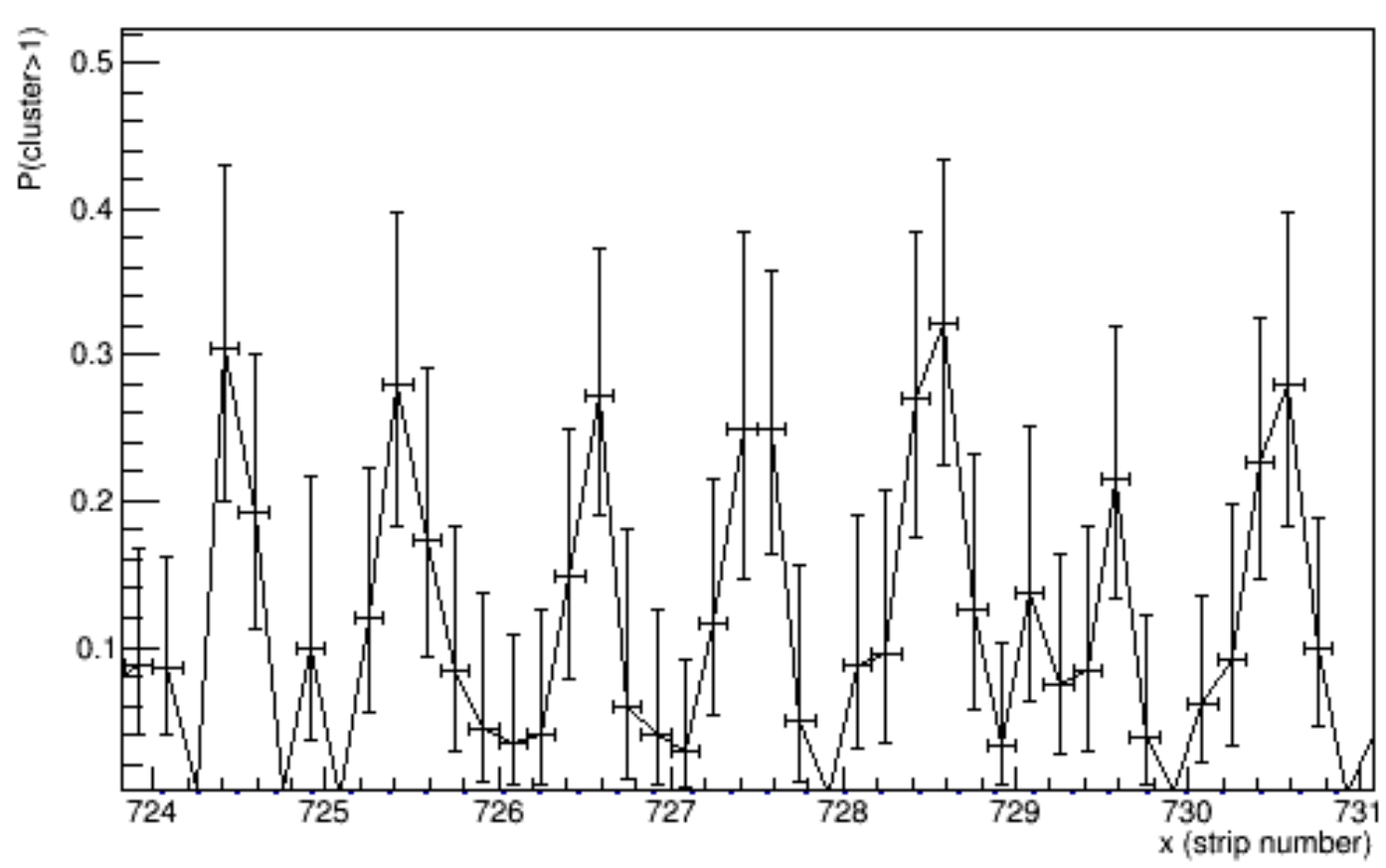

Figure 4.3: Zoom of the cluster size profile along the x direction.

If the cluster size plots show some unexpectedly low cluster size, the reason probably lies in a problem in the transformation from local coordinate system to pixel coordinate system. Usually, this is due to a wrong value of the parameter *beam_offset* in the config file.

To identify any desynchronization during a run, the plot with efficiency as a function of the event number is used. An example of such plot is shown in Figure 4.4, where some blocks of events show an anomalously low efficiency due to desynchronization.

[14]This quantity is roughly similar to Cluster Size - 1

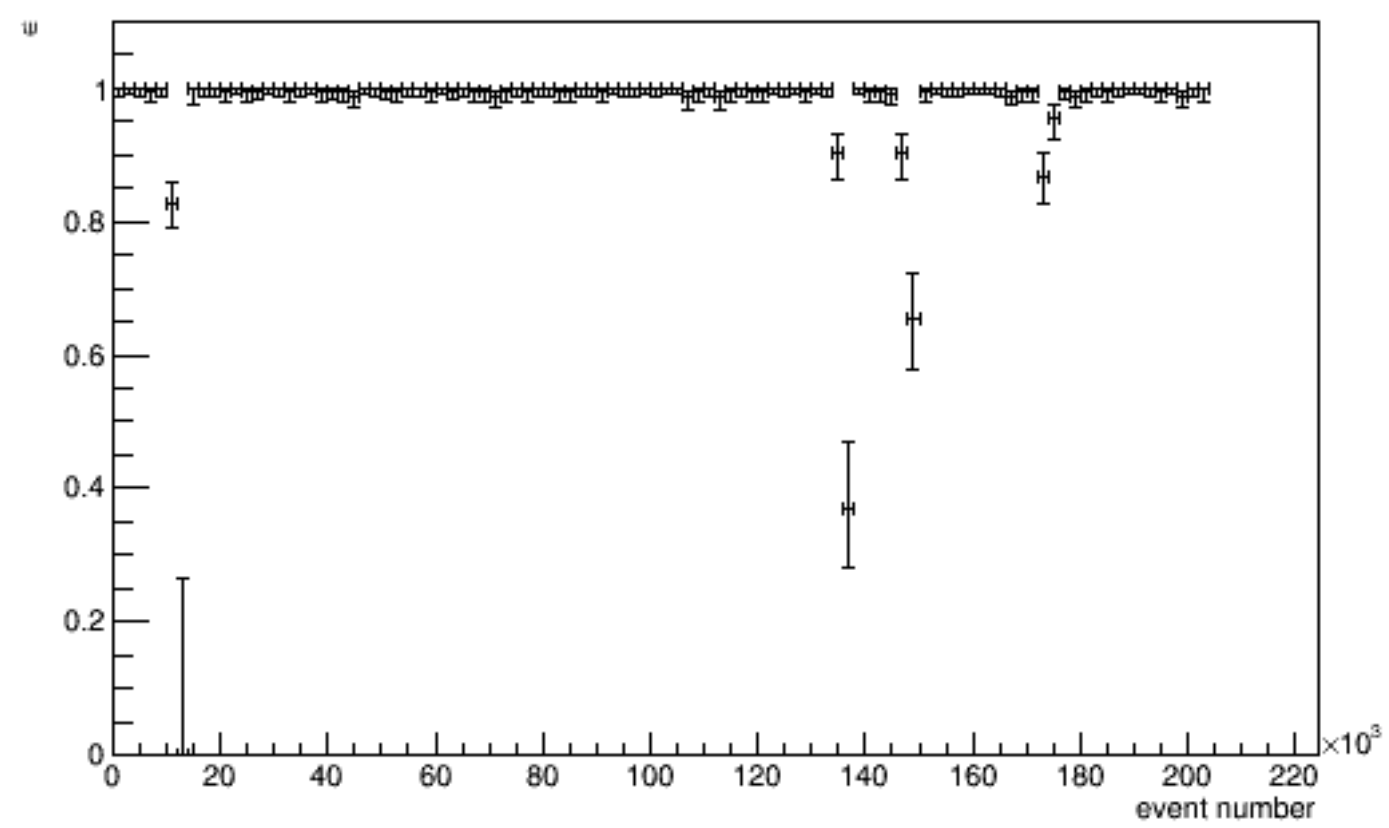

Figure 4.4: Efficiency as a function of the event number. A few desynchronizations can be seen in the form of blocks of events with low efficiency.

Another very important plot is the efficiency vs. time plot. Using this plot at a threshold close to the median charge ($\epsilon \approx 50\%$) it is possible to identify the time window to be used. An example of such a plot is shown in Figure 4.5.

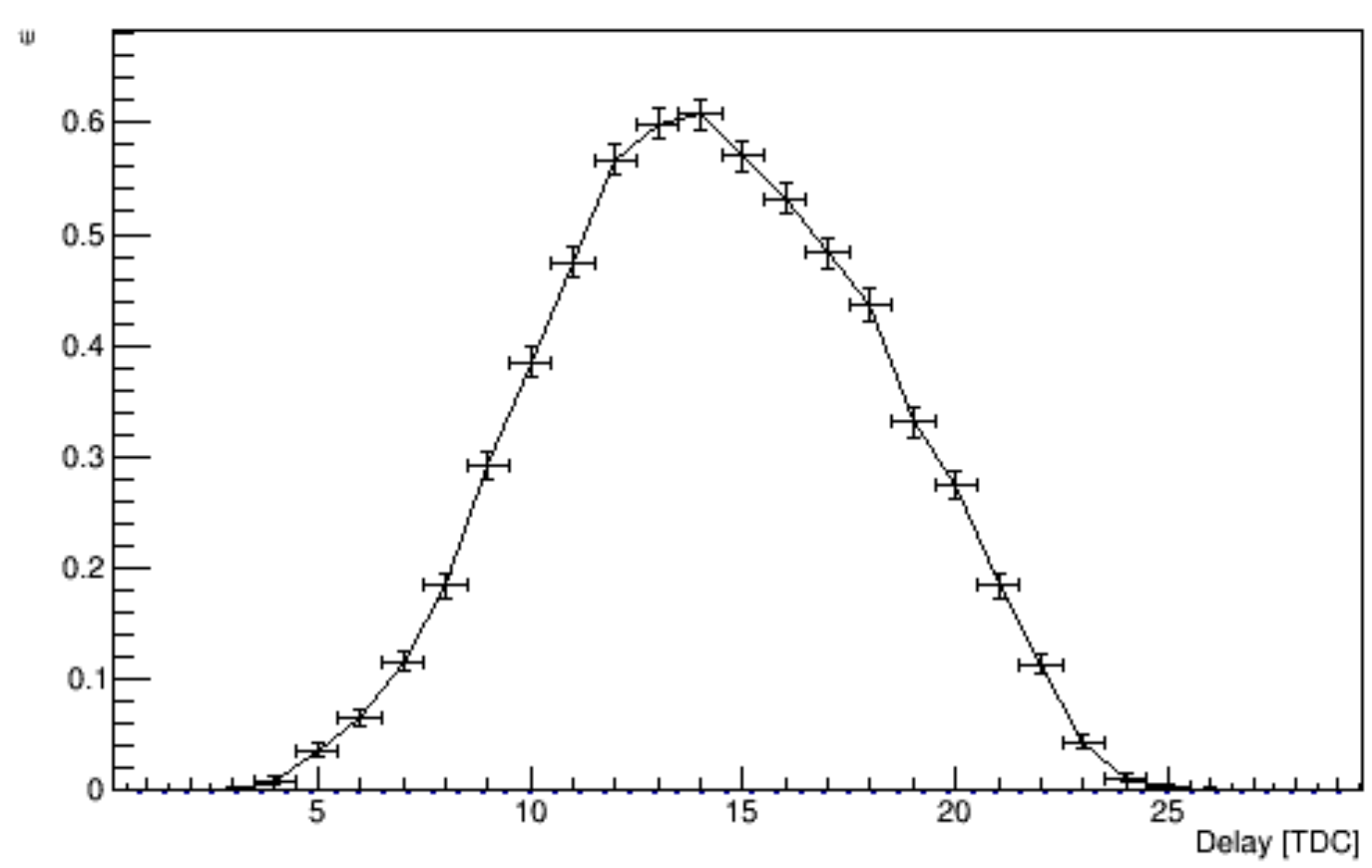

Figure 4.5: Efficiency as a function of the delay for a run at a threshold close to the median charge. The time window should be placed around the peak.

Finally, another useful functionality of this code is the debug mode. It is possible to use it activating the flag –debug:

*./make_testbeam_analysis.sh --config settings.config --input ntuple.root --debug*

In this way, the values contained in some of the TTrees will be printed out in the terminal for a few events.

## 4.2 Second Analysis Code: S-Curves

The second analysis code uses the .root files created by the first code as an input and creates the s-curves. It can be run with the command:

*python plot_efficiency.py*

The input files and their thresholds in DAC are defined in the file input_files.py:

*# thresholds*
*thresholds = [30,45,60,75,85]*
*# input files*
*files = ["TestbeamAnalysis_2767_pos1_30_500.root",*
*"TestbeamAnalysis_2768_pos1_45_500.root",*
*"TestbeamAnalysis_2769_2770_pos1_60_500.root",*
*"TestbeamAnalysis_2771_2772_2774_pos1_75_500.root",*
*"TestbeamAnalysis_2776_2777_pos1_85_500.root"]*

The conversion from DAC to mV, and then from mV to fC are performed by the file plot_tools.py. To convert from DAC to mV, the table in the file ABC130_thrCal.txt is used. For the step from mV to fC the three calibration parameters obtained by the response curve are defined in the form:

*pp=[1125.98 ,3.34, -523.09]*

Finally, in plot_efficiency.py it is possible to set the time window. The time window set in the first code will be applied only on the plots generated by it, but not to the data passed to the second code. The time window must be set in the lines:

*if event.delay >= 20 and event.delay <=25:*

This code will produce several plots regarding efficiency and cluster size. It also creates the desired s-curves, as shown in Figure 4.6, that usually are the primary aim of the ITk Strip test beams.

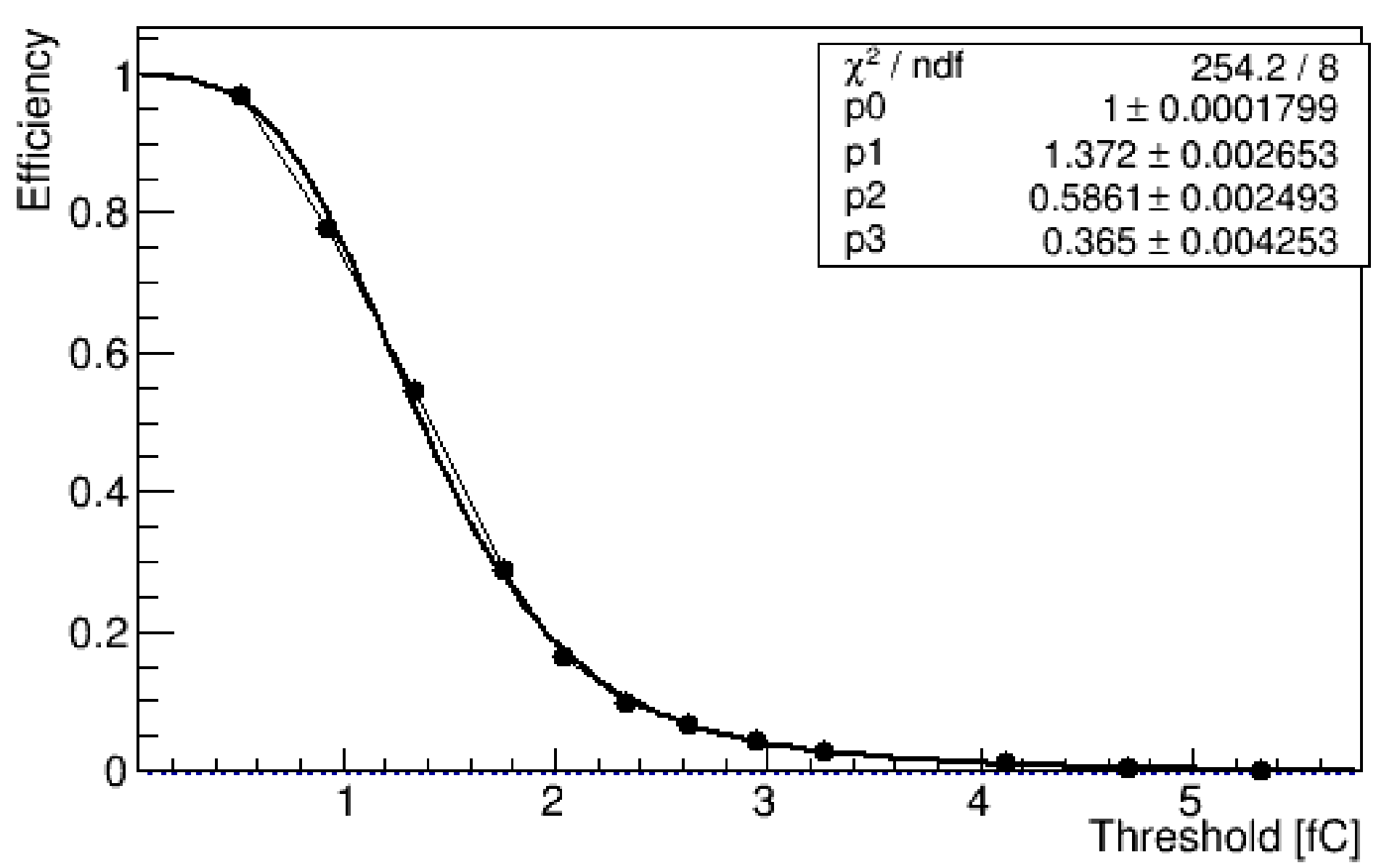

Figure 4.6: Final s-curve created by the second analysis code.

# 5 Appendix

## 5.1 Coordinate Systems

Three different coordinate systems are used in EUTelescope. This is a crucial aspect for the tracking and their knowledge can improve the understanding of all the reconstruction and analysis steps.

The first coordinate system is the *pixel* coordinate system. This is a two-dimensional system and is the one used by EUDAQ. Its definition does not depend on the geometry. The coordinate of the hits is given by the x and y numbers of the pixel (or strip) that has been hit. Figure 5.1 shows an example of a pixel coordinate system.

The second coordinate system is the *local* coordinate system. This is a two-dimensional system as well, and it is the most used in the final track analysis. The center of this coordinate system is the geometrical center of the plane and the x and y positions are defined in mm, and not in pixel numbers. Figure 5.2 shows an example of a local coordinate system. In order to pass from the pixel coordinate system to the local coordinate system, only some intrinsic geometrical properties of the planes are needed: the sensor size and the x and y pitches.
In the ITk Strip EUTelescope version, another parameter is added to the transformation between the pixel and the local coordinate system. This parameter is *localdistDUT* and is defined in the hitmaker section of the config file. This parameter is a shift to the local coordinate system and it allows to set the center of the axis close to the center of the beam spot, instead of the geometrical center of the plane. This facilitates the correlator process. It is necessary to take this shift into account in the post-reconstruction analysis if a transformation between local and pixel coordinate systems is needed.

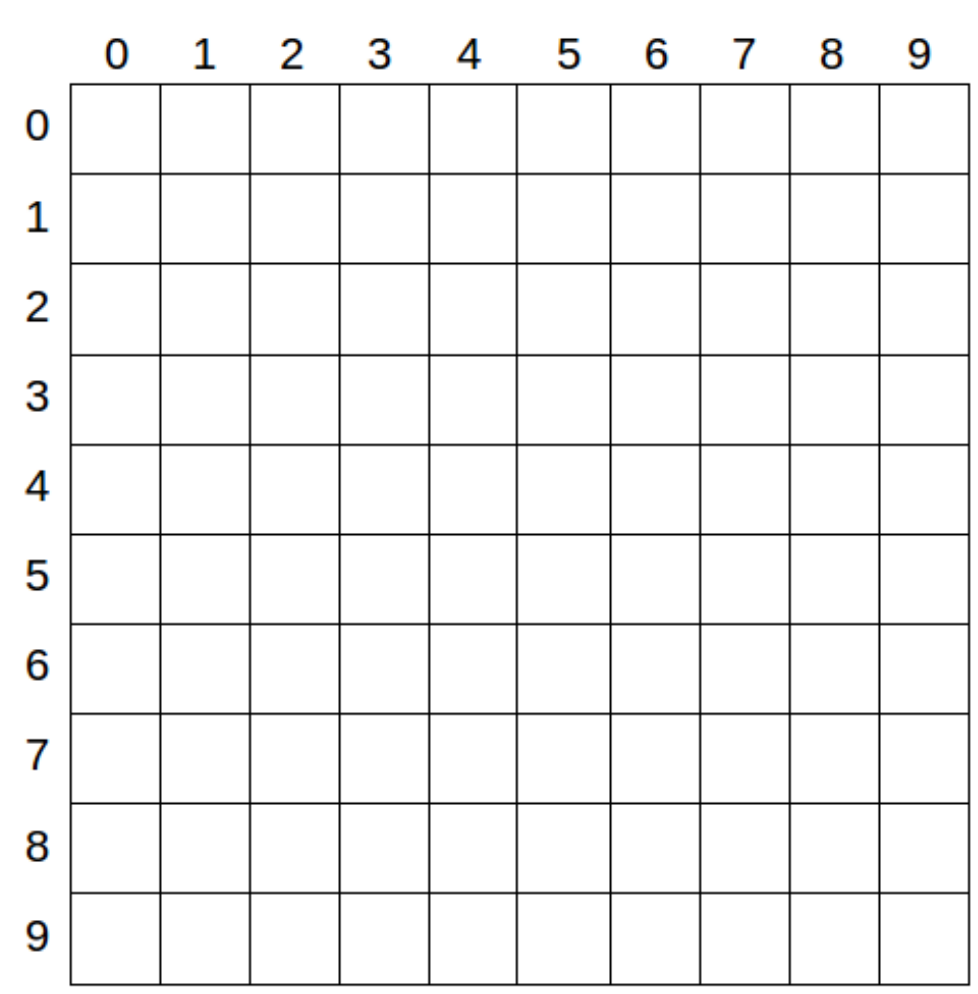

Figure 5.1: Pixel coordinate system

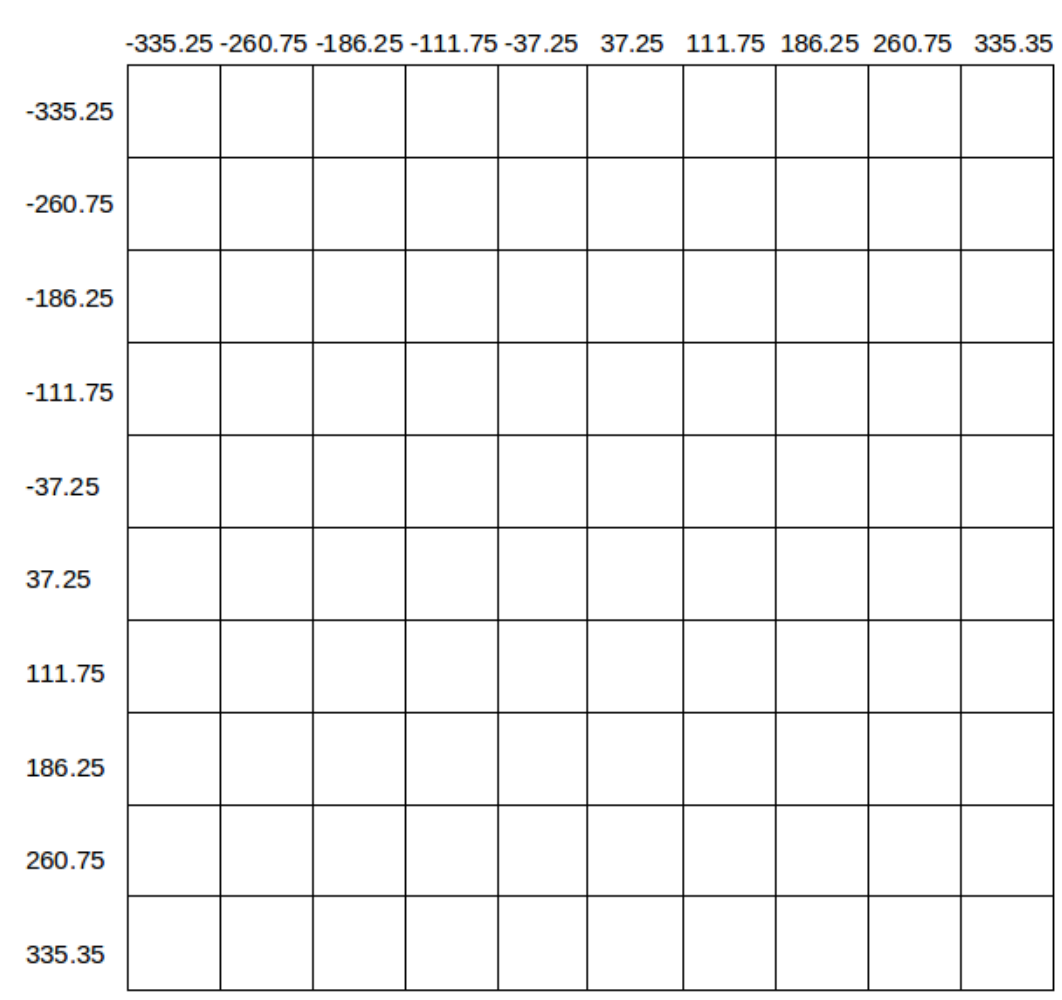

Figure 5.2: Local coordinate system

The last coordinate system is the *global* coordinate system (or telescope coordinate system). This is a three-dimensional system and is the one where the alignment and the tracking are performed. The x and y axes of the global coordinate system are defined by the x and y axes of the first MIMOSA plane. The z axis is the axis that connects the center of the first MIMOSA plane to the center of the last one. To pass from the local coordinate system to the global coordinate system, the x,y and z positions and the xy,xz and yz rotations of the plane are needed. The alignment step provides the correct parameters to transform from the local to the global coordinate system. Despite having a primary role in the reconstruction, this coordinate system is generally not used in the post-reconstruction analysis.

# Acknowledgements

We thank Andrew Blue, Robert Francis Hunter, Graham Greig, Francesco Guescini and Michaela Queitsch-Maitland for the comments that improved this guide and Yi Liu for his huge help with EUDAQ and its conversion process.